\documentclass[pra,twocolumn,preprintnumbers,amsmath,amssymb]{revtex4}
\usepackage{amsmath}
\usepackage{epsfig}
\usepackage{subfigure}
\usepackage{graphicx}
\usepackage{units}
\usepackage{epstopdf}
\usepackage{color}
\usepackage{longtable}

\newcommand{\ketbra}[2]{\mbox{$|#1\rangle\langle #2|$}}
\newcommand{\op}[1]{\mbox{\boldmath $\hat{#1}$}}

\newcommand{\ket}[1]{\vert#1\rangle}
\newcommand{\bra}[1]{\langle#1\vert}

\newcommand{\E}{{\mathcal E}}
\newcommand{\F}{{\mathcal F}}
\DeclareMathAlphabet{\mathpzc}{OT1}{pzc}{m}{it}
\newcommand{\f}{\mathpzc{f}}
\newcommand{\e}{\mathrm{e}}

\begin{document}

\title{Heralded processes on continuous-variable spaces as quantum maps} 
\author{Franck Ferreyrol$^{1,2}$, Nicol\`o Spagnolo$^{1,3}$, R\'emi Blandino$^1$, Marco Barbieri$^{1,4}$ and Rosa Tualle-Brouri$^{1,5}$}
\affiliation{$^1$Laboratoire Charles Fabry, Institut d'Optique, Universit\'e Paris-Sud and CNRS, F-91127, Palaiseau, France}
\affiliation{$^2$Centre for Quantum Dynamics and Centre for Quantum Computation and Communication Technology, Griffith University, Brisbane 4111, Australia}
\affiliation{$^3$Dipartimento di Fisica, Sapienza Universit\`a di Roma, piazzale Aldo Moro 5, I-00185 Rome, Italy}
\affiliation{$^4$Clarendon Laboratory, Department of Physics, University of Oxford, OX1 3PU, United Kingdom}
\affiliation{$^5$Institut Universitaire de France, 103 boulevard St. Michel, 75005, Paris, France}

\begin{abstract}
Conditional evolution is crucial for generating non-Gaussian resources for quantum information tasks in the continuous variable scenario.
However, tools are lacking for a convenient representation of heralded processes in terms of quantum maps for continuous variable states, in the same way as Wigner functions are able to give a compact description of the quantum state.
Here we propose and study such a representation, based on the introduction of a suitable transfer function to describe the action of a quantum operation on the Wigner function.
We also reconstruct the maps of two relevant examples of conditional process, that is, noiseless amplification and photon addition, by combining experimental data and a detailed physical model. This analysis allows to fully characterize the effect of experimental imperfections in their implementations.
\end{abstract}

\maketitle

\section{Introduction}

Quantum mechanics is a probabilistic theory. The quantum description of any experiment is based on probability amplitudes --  quantum states and measurements -- and transformations of such amplitudes -- quantum processes. For each of these objects there exist techniques to obtain the corresponding mathematical tool from measured data: state tomography \cite{Leo97,Jam01}, process tomography \cite{OBr04,Lob08}, and detector tomography \cite{Lun08}. There exist constraints necessary to attribute a near physical meaning to abstract mathematics: for instance, a map acting on density matrices space corresponding to a physical process is normally completely positive (CP). This amounts to say that it must send physical states into physical states regardless of observing the system by itself or as a part of a larger ensemble to which it is de-coupled \cite{Kra83}. Most of studied maps preserve the norm of the state, but there exist notable exceptions: non-trace preserving operations arise whenever a measurement on the system is involved.

An interesting class of such processes involves heralding: the evolution of a system is considered conditionally on the outcome of a measurement on part of the system itself [\ref{fig:conceptual}]. While this is a legitimate operation in classical physics,  in quantum mechanics non-standard behaviour may arise: this is the case, for instance, of anomalous weak values \cite{Aha88}. In the context of optical quantum information, conditional evolution has found several applications for simulating strong nonlinearities at the few-photon level. This approach has allowed to build two-qubit \cite{OBr04} and three-qubit quantum logic gates \cite{Lan08} and to generate quantum states with non-Gaussian Wigner function \cite{Lvo01, Wen04, Zav04, Our06, Our07}. Such an evolution is able to induce non-Gaussian transformations effective in overcoming existing no-go theorems valid for purely Gaussian resources \cite{Eis02, Fiu02, Nis09}. Successful applications of such processes to communication tasks have been demonstrated in several experiments \cite{Our07a, Tak10, Fer10, Xia10}.

The experimental investigation is relatively at an early stage: so far quantum process tomography of non trace-preserving maps has been presently implemented only in a reduced two-qubits Hilbert space \cite{Kie05,Bon10}. 
Here we show, by a detailed physical model, the description of two conditioned processes which are relevant to continuous-variable state manipulation: the noiseless amplifier \cite{Fer10,Xia10} and the single-photon addition \cite{Zav04,Bar10}. We can derive the expression of the map in the well known tensor form, and,  as a step further, we illustrate a transfer function formalism, which allows to describe quantum process directly in the Wigner representation. This will stimulate to deepen investigations in this area and to develop more sophisticated analytic tools.

\section{Quantum maps}

Any transformation acting on states needs to satisfy some physically-motivated mathematical constraints. In the simplest case, a closed system, the evolution of a quantum states is described by a unitary operator $\op U$, which transforms the input state as $\rho'{=}\op{U}\rho\op{U}^\dag$. More generally, the system will be able to interact with the environment and a representation in terms of a unitary won't be sufficient to describe this scenario; however, some essential features are retained, in particular the output state must be obtained from a linear transformation of the input. The proper formalism then adopts a generic linear map $\E$ such that $\rho'=\E(\rho)$. 
Similarly to the previous expression, this map can be  decomposed in the incoherent application of a set of Kraus operators $\{\op{E}_i\}$ \cite{Kra83}:
\begin{equation}
\label{Krauss}
\E(\rho)=\sum_{i}\op{E}_{i}\rho \op{E}_{i}^\dag.
\end{equation}
One can note that this expression is similar to the formalism used for positive-operator valued measure (POVM) since a generic transformation can be seen as the application of a unitary operation on a system composed by the input state and the environment, followed by a measurement of the environment for which we do not know the outcome.
Another expression, more convenient for data visualisation, uses a tensor $\{\E_{l,k}^{n,m}\}$:
\begin{equation}
\label{rho2rho}
[\E(\varrho)]_{l,k}=\sum_{n,m}\E_{l,k}^{n,m}\varrho_{n,m}.
\end{equation}
Where the elements of the tensor are given by $\E_{l,k}^{n,m}=\sum_i\bra{l}\op{E}_{i} \ketbra{n}{m}\op{E}_{i}^\dag \ket{k}$

These maps can not be completely arbitrary: an essential requirement is that they lead physical states in physical states. Therefore, these maps have to send positive operators into positive operators, and, for deterministic processes, require to preserve the trace; so they directly give a physical density matrix without need of other operation.
Furthermore, in the majority of the cases, we also demand complete positiveness: this amounts to say that the evolution must remain physical when the system is entangled with a second object. 

A somehow different context arise when considering a conditional process : it implies a non-linear evolution of the state due to the re-normalisation operation.
Indeed, these processes often aim to approximate a non-unitary linear operator $\op C$ and transform a pure state $\ket{\alpha}$ into  $\sqrt{N(\alpha )}\op C\ket{\alpha}$, where $N(\alpha)$ is the normalisation factor, which might present a complex dependence on the state. And, even if $\op C$ is actually linear, this linearity is shadowed if we consider the physical inputs, as  ${\sqrt{N(\alpha )}\op C\ket{\alpha}+\sqrt{N(\beta )}\op C\ket{\beta}\neq\sqrt{N(\alpha+\beta)}\op C(\ket{\alpha}+\ket{\beta})}$. As a result, in order to represent it as a linear quantum map one should use a trace non-preserving map and keep the normalisation step for the result of the process. 
Moreover, a map should include one more information to describe a conditional process which is the success probability. 

\begin{figure}[htb!]
\includegraphics[width=0.48\textwidth]{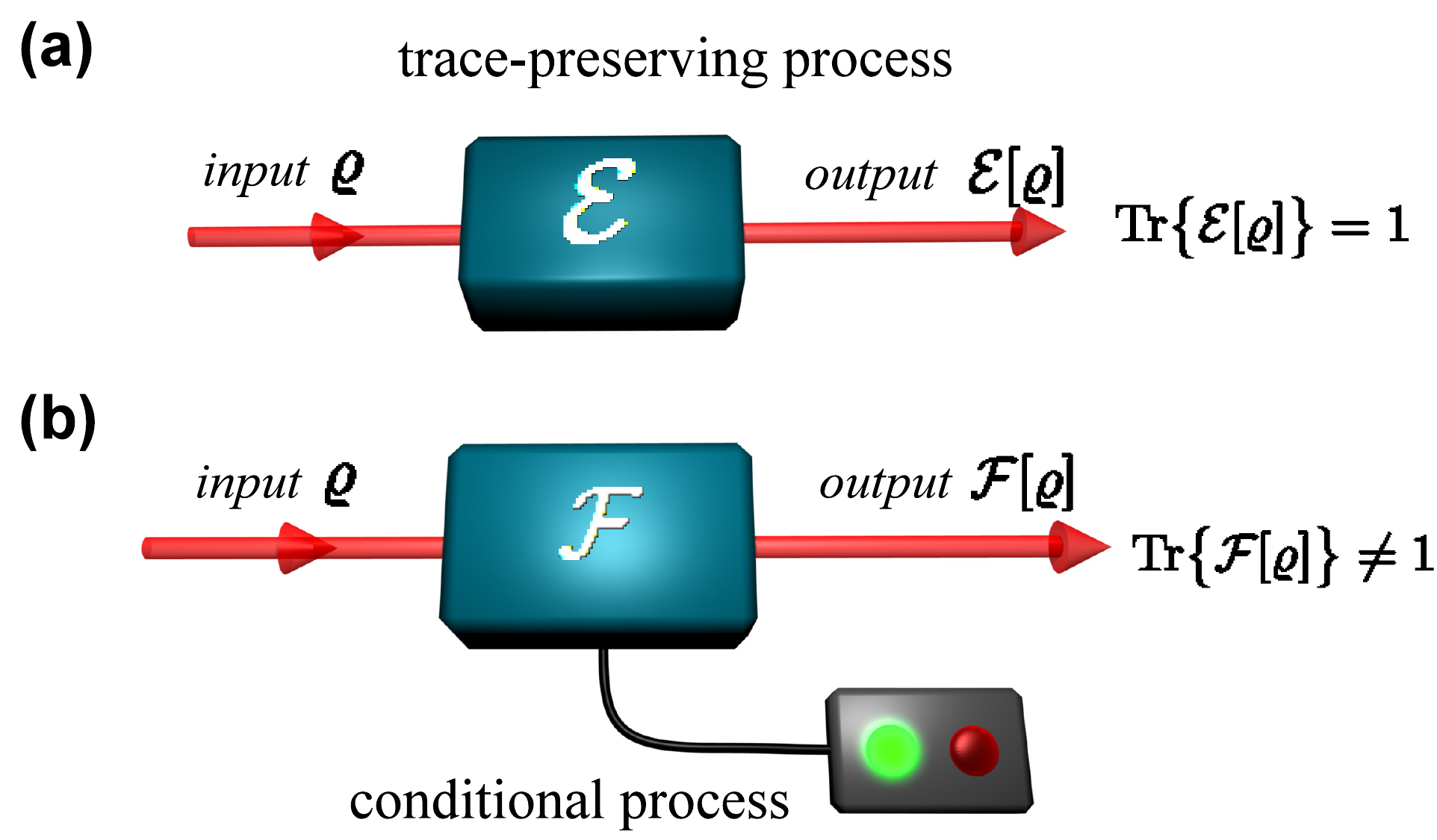}
\caption{(a) Trace-preserving quantum operation. The input state $\varrho$ is transformed by the quantum channel $\mathcal{E}$ in the output state $\mathcal{E}[\varrho]$.
The trace of the input state is preserved by the channel: $\mathrm{Tr}\big\{ \mathcal{E}[\varrho]\big\} = 1$. (b) Heralded quantum operation. The input state 
$\varrho$ is transformed by the quantum channel $\mathcal{F}$ in the output state $\mathcal{F}[\varrho]$ upon realization of a conditional event. The trace of the input 
state is in general not preserved by the channel: $\mathrm{Tr}\big\{ \mathcal{F}[\varrho]\big\} \neq 1$.}
\label{fig:conceptual}
\end{figure}

In fact, a conditional evolution acts as following: 
the system evolves through a probabilistic device, and we only accept those runs when a successful event is flagged (Fig. \ref{fig:conceptual}). Clearly, the overall process including both successes and failures can be modelled by a deterministic quantum map, and in terms of Kraus decomposition (eq. \ref{rho2rho}) passing from the overall process to the conditional one, consist to keep only the subset of the Kraus operators corresponding to the result of the measure used for heralding. Thus we easily see that the trace of the output state gives our additional information: the success probability, and a quantum map only need to be trace non increasing to correspond to a physical operation \cite{Nielsen_Chuang}.
Nevertheless, such definition can then be extended to more general processes, as far as the transformation remain linear in the quantum state. In particular it can be used with trace increasing maps, which, even if they are not physical, are very common in theoretical quantum  physics and include most of the non-unitary operators that are approximated by conditional processes.

The method explained above gives a neat picture of the process for discrete-variable systems: for instance, one can recognise almost at glance the behaviour of a qubit process by inspecting the corresponding tensor. This is more complex when dealing with continuous variable systems, when often  looking at the states as Wigner quasi-distribution in the phase space can convey information in a more compact and effective way. Therefore, a method to represent quantum processes in the Wigner representation would be highly desirable. Such an object has been proposed for the unitary processes \cite{Ber79, Coh83} and for Gaussian operations \cite{Fiu02}; here we give an explicit extension of these results to the case of a generic map $\E$.

For this purpose, we can reason in analogy with the probability distribution ${\cal P}(x,p)$ for physical position and momentum of a classical particle. The action of a Markovian process will modify such distribution via a transfer function $\f(x',p',x,p)$ describing the odds that a particle initially in the position $(x,p)$ will eventually end in $(x',p')$. The distribution ${\cal P}'(x',p')$ of the coordinates at the end of the process will result from the sum of all these elementary displacements: 
\begin{equation}
\label{eq:transfer_class}
{\cal P}'(x',p')=\int dx\,dp\,{\cal P}(x,p)\f(x',p',x,p).
\end{equation}
Hence, we would like to maintain this structure for quantum processes as well by introducing a suitable transfer function $\f_{\E}(x',p',x,p)$ by which the input Wigner function $W(x,p)$ can be turned into the output $W'(x',p')$ by the integral transform:
\begin{equation}
\label{eq:transfer_def}
W'(x',p')=\int dx\,dp\,W(x,p)\f_\E(x',p',x,p).
\end{equation}

In order to see that this is actually the case, we start from the case where only one Kraus operator $\op{E}_{i}$ is present; the general result can be obtain by linearity. The Wigner function of the output state then reads:
\begin{equation}
W'(x',p')=\frac{1}{2\pi}\int d\nu\, e^{i\nu p'} \bra{x'-\frac{\nu}{2}}\op{E}_{i}\rho \op{E}_{i}^\dag\ket{x'+\frac{\nu}{2}}.
\end{equation}
We invoke the completeness relation so to obtain
\begin{widetext}
\begin{equation}
W'(x',p')=\frac{1}{2\pi}\int dp\,d\nu\,ds\,dt\, e^{i\nu p'}e^{i(s-t)p} W(\frac{s+t}{2},p)\bra{x'-\frac{\nu}{2}}\op{E}_{i}\ket{s}\bra{t} \op{E}_{i}^\dag \ket{x'+\frac{\nu}{2}}.
\end{equation}
and then, by a variable substitution, the expression for the transfer function associated to the operator $\op{E}_i$:
\begin{equation}
f_{i}(x',p',x,p)=\frac{1}{2\pi}\int d\mu\,d\nu\, e^{i\nu p'}e^{i\mu p} \bra{x'-\frac{\nu}{2}}\op{E}_{i}\ket{x+\frac{\mu}{2}}\bra{x-\frac{\mu}{2}}\op{E}_{i}^\dag \ket{x'+\frac{\nu}{2}},
\label{eq:kraus-func}
\end{equation}
\end{widetext}
which implies
\begin{eqnarray}
	\int f_i(x',p',x,p) dx' dp'  &=& W_{\op{E}_{i}^\dag\op{E}_{i}}(x,p)\\
	\int f_i(x',p',x,p) dx dp & =& W_{\op{E}_{i}\op{E}_{i}^\dag}(x',p')
\end{eqnarray}
Based on the remark that each $\op{E}_i$ acts independently, the transfer function associated to a generic process reads:
\begin{equation}
f_{\E}(x',p',x,p)=\sum_i f_{i}(x',p',x,p)
\label{eq:kraus-sum}
\end{equation}
where each function $f_i$ corresponds to a Kraus operator with the relation given by Eq. (\ref{eq:kraus-func}).
 It can be checked that using this formula as the definition, and using the relation between $\E_{l,k}^{n,m}$ and $\op{E}_{i}$, one arrives to the original definition. The transfer function might be a distribution, but from Eq. (\ref{eq:kraus-func}) it appears that it is always real. Also, normalisation enforces that
\begin{eqnarray}
	\int f_i(x',p',x,p) dx' dp' dx dp 	& = & \mathrm{Tr}(\op{E}_{i}^\dag\op{E}_{i}) \\
							& = & \mathrm{Tr}(\op{E}_{i}\op{E}_{i}^\dag)
\end{eqnarray}
Those properties can easily be extended to the whole transfer function. In particular, in the case of a deterministic map we have
\begin{equation}
	\int f_i(x',p',x,p) dx' dp'  = 1 ,
\end{equation}
whereas for a non-deterministic process the integral of $W_{\op{E}_{i}^\dag\op{E}_{i}}(x,p)W(x,p)$ gives the success probability.

We also can express the transfer function in terms of the process tensor, starting with the Eq.  \eqref{rho2rho} rewritten in the form
\begin{equation}
\E(\varrho)=\sum_{n,m}\sum_{l,k}\E_{l,k}^{n,m}\mathrm{Tr}\left(\varrho\cdot\ketbra{n}{m}\right)\ketbra{l}{k}.
\end{equation}
We can use the properties of the Wigner functions to evaluate the expectation value $\mathrm{Tr}\left(\varrho\cdot\ketbra{n}{m}\right)$, and derive:
\begin{widetext}
\begin{equation}
W'(x',p')=2 \pi \int dx\,dp\, W(x,p)\left(\sum_{m,n}\sum_{l,k}\E_{l,k}^{n,m}W_{\ketbra{n}{m}}(x,p)W_{\ketbra{l}{k}}(x',p')\right),
\end{equation}
\end{widetext}
where $W_{\ketbra{l}{k}}(x,p)$ is the Wigner representation of $\ketbra{l}{k}$. Therefore, the quantum operation $\E$ is conveniently represented by the transfer function
\begin{equation}
\begin{aligned}
\f_\E(x',p',x,p)&=2 \pi \sum_{n,m}\sum_{l,k}\E_{l,k}^{n,m}W_{\ketbra{n}{m}}(x,p) \times\\ &\times W_{\ketbra{l}{k}}(x',p').
\end{aligned}
\end{equation}

The quantum transfer function still bares resemblance with Markovian processes. This can be seen by inspecting what happens when chaining two processes $\E{=}\E_1\otimes\E_2$. Under these circumstances, we obtain the complete transfer function
as: \begin{equation}
\f_\E(x',p',x,p)=\int dx''\,dp''\,\f_{\E_2}(x',p',x'',p'')\f_{\E_1}(x'',p'',x,p).
\label{eq:comb-map}
\end{equation}
which is similar to the Chapman-Kolmogorov equation for Markovian processes \cite{Coh83}.
 This reinforces the view that from a classical viewpoint, $\f_\E(x',p',x,p)$ should be interpreted as a  a transition probability from $\{x,p\}$ to $\{x',p'\}$. The analogy can not be extended further in the quantum domain: in the following, we will illustrate a case where $\f_\E(x',p',x,p)$ can actually take negative values. However, we will also show how the temptation of establishing a direct quantitative connection between nonclassicality and the negative values of $\f_\E$ has to be resisted.

\section{Determination of the transfer function} 

In order to determine the expression of the transfer function, one could first decompose it in a sum of transfer function corresponding to the different heralding events, in a similar way as eq. \ref{eq:kraus-sum}. Then, each of those transfer function can be constructed by composing, with the use of Eq. \ref{eq:comb-map}, of some basic transfer functions corresponding to the different elements of the process.The basic transfer functions can have the same number of input and output mode, corresponding to a basic transformation, only output modes, corresponding to the introduction of an ancilla state, or only input modes, corresponding to a measurement.

The basic transformations can be determined by Eq. \ref{eq:kraus-func} or by simple considerations. In particular all transformation that can be expressed as a coordinate transformation have a transfer function composed of Dirac distributions which directly come from the coordinate transformation.
An other interesting basic transformation is the one given by a coordinate transformation $M_i$ ($i=x,p$) for each quadrature of the input mode and a quadrature of a vacuum ancilla, followed by a partial trace on the ancilla. If the transformation matrix is
\begin{eqnarray}
M_i &=&
\left(
\begin{matrix}
\mu_i && \nu_i\\
\varepsilon\nu_i && \mu_i
\end{matrix}
\right),\\
\mathrm{det}(M_i) & = & 1,
\end{eqnarray}
where $\varepsilon=\pm1$, then the transfer function of this operation is:
\begin{widetext}
\begin{equation}
f(x',p',x,p)=\frac{\nu_x\nu_p}{\pi}\exp\left(-\left(\frac{x'-\mu_x x}{\nu_x}\right)^2\right)\exp\left(-\left(\frac{p'-\mu_p p}{\nu_p}\right)^2\right).
\end{equation}
\end{widetext}
Table \ref{tbl:basic_transfo} shows some basics transfer functions obtained by those considerations with their tensor form.

\begin{table*}[htbp!]
\begin{ruledtabular}
\caption{Tensor process and transfer functions of some basic transformations}
\begin{tabular}{|c|c|c|}
\hline
Transformation & tensor & transfer function \\
\hline
Identity & $\delta_{k,n}\delta_{l,m}$ & $\delta(x-x')\delta(p-p')$ \\
\hline
Phase rotation & $\e^{\theta(k-l)}\delta_{k,n}\delta_{l,m}$ & $\delta(x - \cos\theta x'+\sin\theta p') \delta(p - \cos\theta p'-\sin\theta x')$ \\
\hline
Displacement & $\sqrt{m!n!l!k!}\e^{|\alpha|^2}\sum_{i=0}^{m}\sum_{j=0}{n}\binom{m}{i}\binom{n}{j}$ & $\delta(x-x'+\sqrt{2}\mathrm{Re}(\alpha)) \delta(p-p'+\sqrt{2}\mathrm{Im}(\alpha))$ \\
 & $\frac{(-1)^{m+n-i-j}}{(l-i)!(k-j)!}\alpha^{l+n-i-j}\bar{\alpha}^{k+m-i-j}$ & \\
\hline
& $\frac{\sqrt{k!l!}}{(m!n!0^{3/2}}\cosh(r)^{k+l-1}\sum_{i=o}^k\sum_{j=0}^l 2^{i+j}$ &  \\
Squeezing  & $\times \frac{\sqrt{(n+k-2i)!(l+m-2j)!}}{\left(\frac{n+k}{2}-i\right)!\left(\frac{m+l}{2}-j\right)!} \left(-\frac{\tanh(r)}{2}\right)^{(m+n+l+k)/2}$ & $\delta(x-\e^{r}x') \delta(p-\e^{-r}p')$ \\
 & $\times \frac{(n+k-i)!}{i!(k-i)!}\frac{(m+l-j)!}{k!(l-j)!} \frac{1+(-1)^{n+k}}{2}\frac{1+(-1)^{m+l}}{2}$ & \\
\hline
 & $\sqrt{\frac{m_1!m_2!n_1!n_2!}{l_1!l_2!k_1!k_2!}}\sum_{i=0}^{l_1}\sum_{j=0}^{k_1}(-1)^{l_1+k_1-i-j}$ & \\
Beam splitter & $\times \binom{l_1}{i}\binom{l_2}{m_1-i} \binom{k_1}{j}\binom{k_2}{n_1-j} $ & $\delta(x_1 - tx_1'+rx_2') \delta(p_1 - tp_1'+rp_2')$\\
 & $\times t^{2i+2j+l_2+k_2-m_1-n_1}  r^{l_1+k_1+m_1+n_1-2i-2j}$ & $ \times \delta(x_2 - tx_2'-rx_1') \delta(p_2-tp_2'-rp_1')$ \\
 & $\times \delta_{m_1+m_2,l_1+l_2}\delta_{n_1+n_2,k_1+k_2}$ & \\
\hline
 & $\sqrt{\frac{m_1!m_2!n_1!n_2!}{l_1!l_2!k_1!k_2!}} \frac{(g-1)^{(m_1+n_1)/2}}{g^{(m_1+n_1+l_2+k_2)/2+1}} $ & \\
Parametric & $\times \delta_{n_2-n_1,k_2-k_1}  \delta_{m_2-m_1,l_2-l_1}$ & $\delta(\sqrt{g}x_1+\sqrt{g-1}x_2-x_1') \delta(\sqrt{g}p_1-\sqrt{g-1}p_2-p_1')$ \\
  down-conversion & $\times \sum_{i=0}^{k_1}\sum_{j=0}^{l_1}  \frac{(n_2+i)!(m_2+j)!}{(n_1-k_1+i)!(m_1-l_1+j)!} $ &  $\times \delta(\sqrt{g}x_2+\sqrt{g-1}x_1-x_2') \delta(\sqrt{g}p_2-\sqrt{g-1}p_1-p_2')$ \\
 & $\times \frac{(-1)^ {i+j} g^{k_1+l_1-i-j} (g-1)^{i+j-(k_1+l_1)/2}}{i!(k_1-i)!j!(l_1-j)!} $ & \\
\hline
Attenuation & $\sqrt{\frac{m!n!}{l!k!}}\frac{\eta^{(l+k)/2}(1-\eta)^{m-l}}{(m-l)!}\delta_{m-l,n-k}$ & $\frac{1}{\pi(1-\eta)}\exp\big(-\frac{(x'-\sqrt{\eta}x)^2}{1-\eta}-\frac{(p'-\sqrt{\eta}p)^2}{1-\eta}\big)$ \\
\hline
Parametric amplification & $\sqrt{\frac{l!k!2^{n+m}}{n!m!2^{l+k}}} \frac{2^{k-m}}{(k-m)!} \frac{g^{k+(m-n)/2}}{(g+1)^{k+n+1}} \delta_{l-n,k-m}$ & $\frac{1}{\pi(g-1)}\exp\big({-\frac{(x'-\sqrt{g}x)^2}{g-1}-\frac{(p'-\sqrt{g}p)^2}{g-1}}\big)$ \\
\hline
\end{tabular}
\end{ruledtabular}
\label{tbl:basic_transfo}
\end{table*}

The basic functions with only output or input modes are even simpler to determinate. Indeed the first ones are exactly the Wigner function of the introduced ancilla, while the second ones are the Wigner function of the projector corresponding to the measure multiplied by a factor $2\pi$. With those considerations, the transfer function can in fact be seen as an extension of the Wigner function.

Finally, the different heralding events are generally the ideal heralding and the faulty ones. Consider for instance a heralded process when conditioning can be faulty in a certain fraction of the total events.  We can call $\F_1$ the correct process, and $\F_2$ the failure. The output $\varrho_{out}$ state of the whole process will be a convex combination of 
\begin{equation}
\varrho_{out} = \xi\frac{\F_1(\varrho_{in})}{P_1} + (1-\xi)\frac{\F_2(\varrho_{in})}{P_2},
\end{equation} 
where $P_{1,2}=\mathrm{Tr}[\F_{1,2}(\varrho_{in})]$ are the success probability used here to normalize the results of both maps. If we notice that $\xi = P_1/(P_1+P_2)$, we can then infer that the transformation which includes both events is given by
\begin{equation}
\varrho_{out} = \frac{\F_1(\varrho_{in})+\F_2(\varrho_{in})}{P_1+P_2}
\end{equation} 
where $P_1+P_2$ is effectively the trace of $\F_1(\varrho_{in})+\F_2(\varrho_{in})$. This amounts to say the correct map is $\F{=}\F_1+\F_2$, provided that we choose $\F_2$ in order to have the correct occurrence probability. This last point is the most crucial, and the determination of the good faulty process $\F_2$ can be complicated. The simplicity of the expression is due to the fact that the maps already contains the success probability for heralded process. Nevertheless we should notice that if the false heralding comes from a noisy mode in the input one should add (at least) a supplementary input mode to the map.

\section{Example 1: the noiseless amplifier}

We now inspect two important quantum processes with the formalism of quantum maps: we will be able to highlight clear signatures of nonclassicality, and observe how they degrade under experimental conditions. As an interesting feature, we will be able to capture such nonclassical aspects at a glance. Our analysis first concerns the noiseless amplifier ($\op{C}=g^{\op{n}}$, where $\op{n}$ is the number operator and $g>1$) \cite{Fer10,Xia10}. We do not adopt a ``black box'' approach, rather a model of the process is used so to arrive to a description in term of generalised maps, also in the case when all the imperfections are taken into account. 

\begin{figure}[htbp]
\centering
\includegraphics[width=0.5\textwidth]{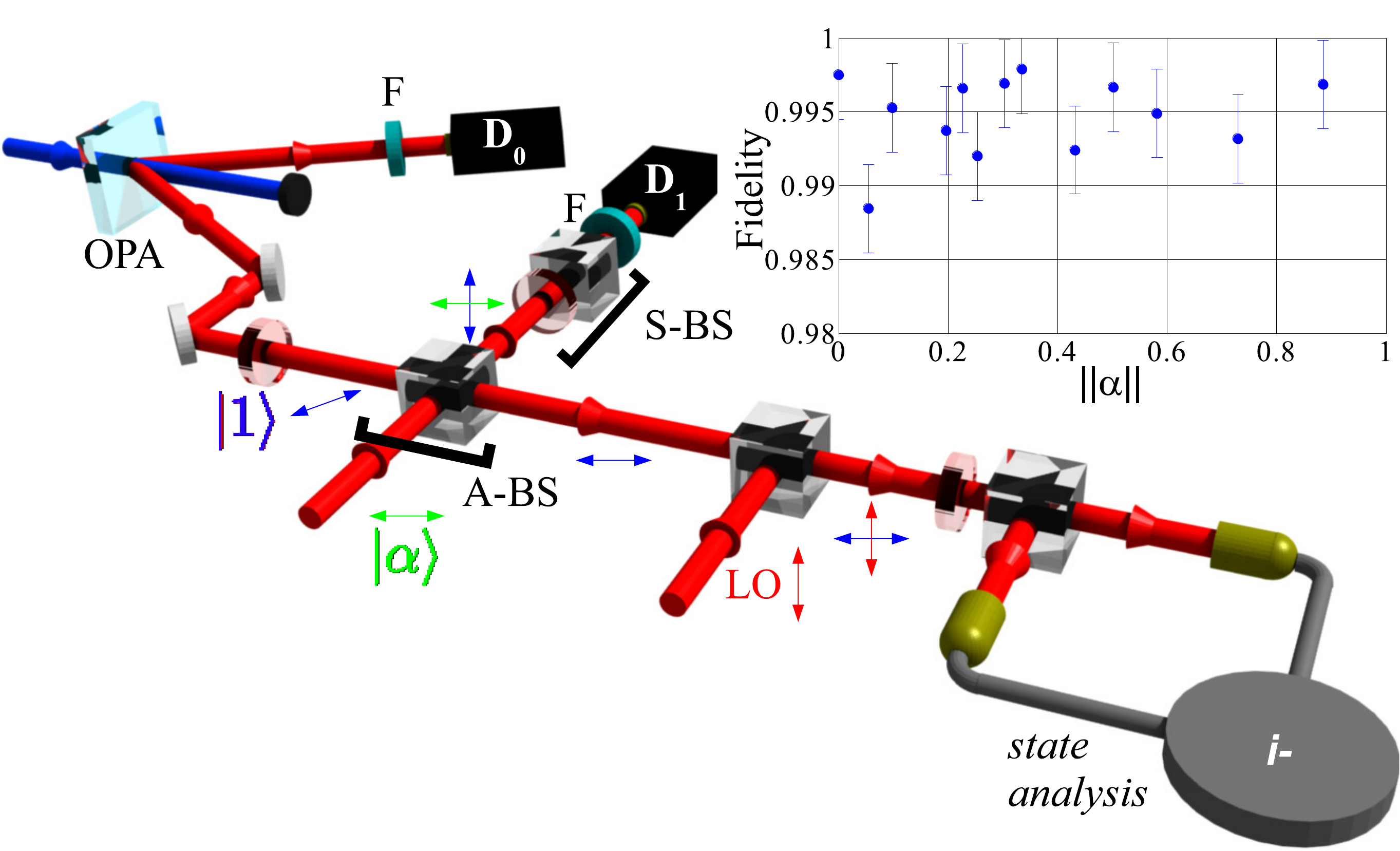}
\caption{Layout of the implementation of the noiseless amplifier \cite{Fer10}. A single photon
is conditionally generated upon detection of a single photon on detector $D_{0}$. After splitting in 
an asymmetric beam-splitter (A-BS), the single photon is mixed with the input coherent state $\vert
\alpha \rangle$ in a symmetric beam-splitter (S-BS). The noiseless amplification process occurs
conditionally to the detection of a photon on detector $D_{1}$. The beam-splitter operations are
performed exploiting polarization (double sided arrows in the figure). Inset: fidelities 
between the experimental density matrices \cite{Fer10} and the prediction of the model.}
\label{fig:actual_noiseless}
\end{figure}

\begin{figure}[htbp]
\centering
\subfigure[]{\includegraphics[width=0.235\textwidth]{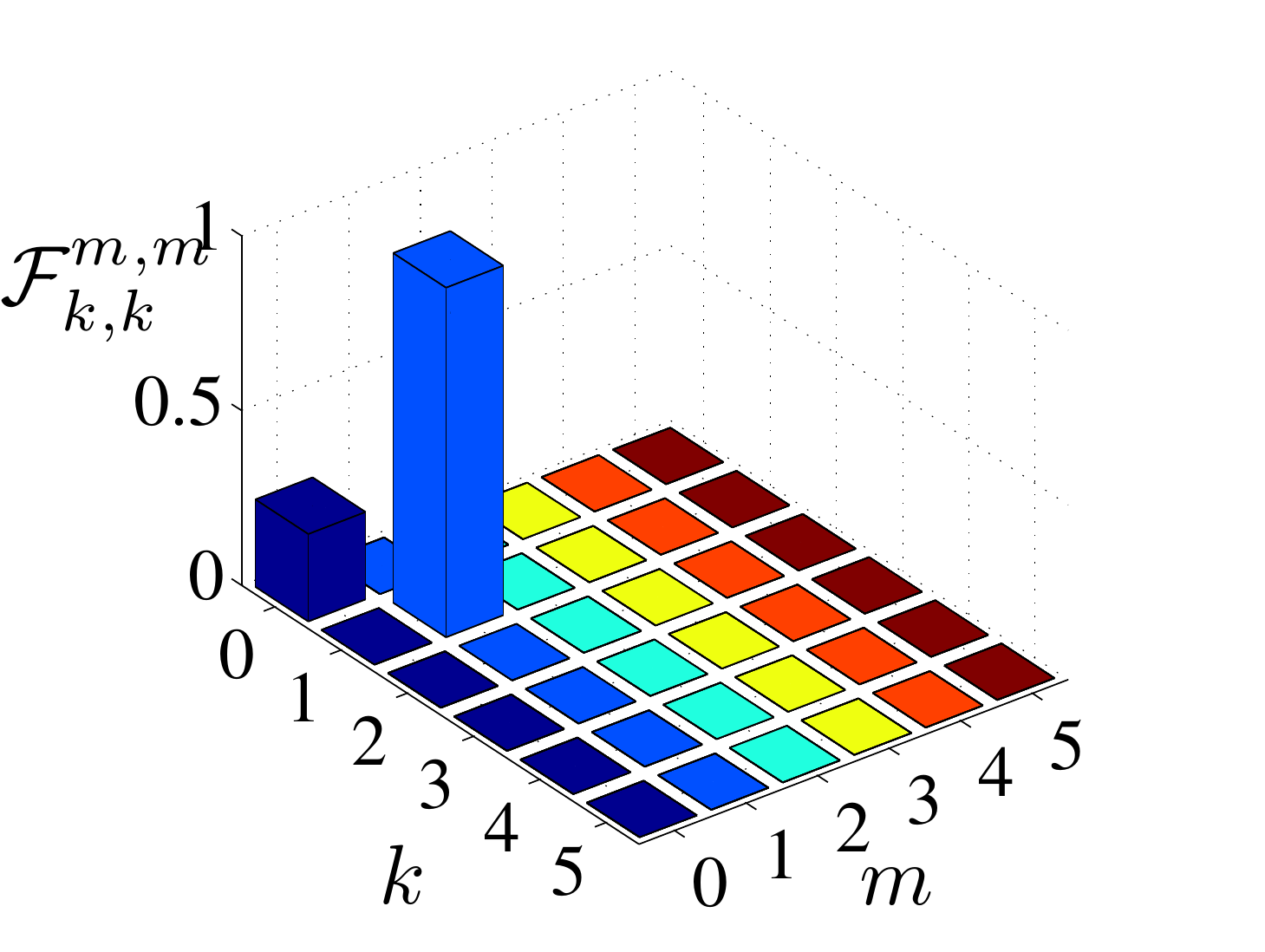}}
\subfigure[]{\includegraphics[width=0.235\textwidth]{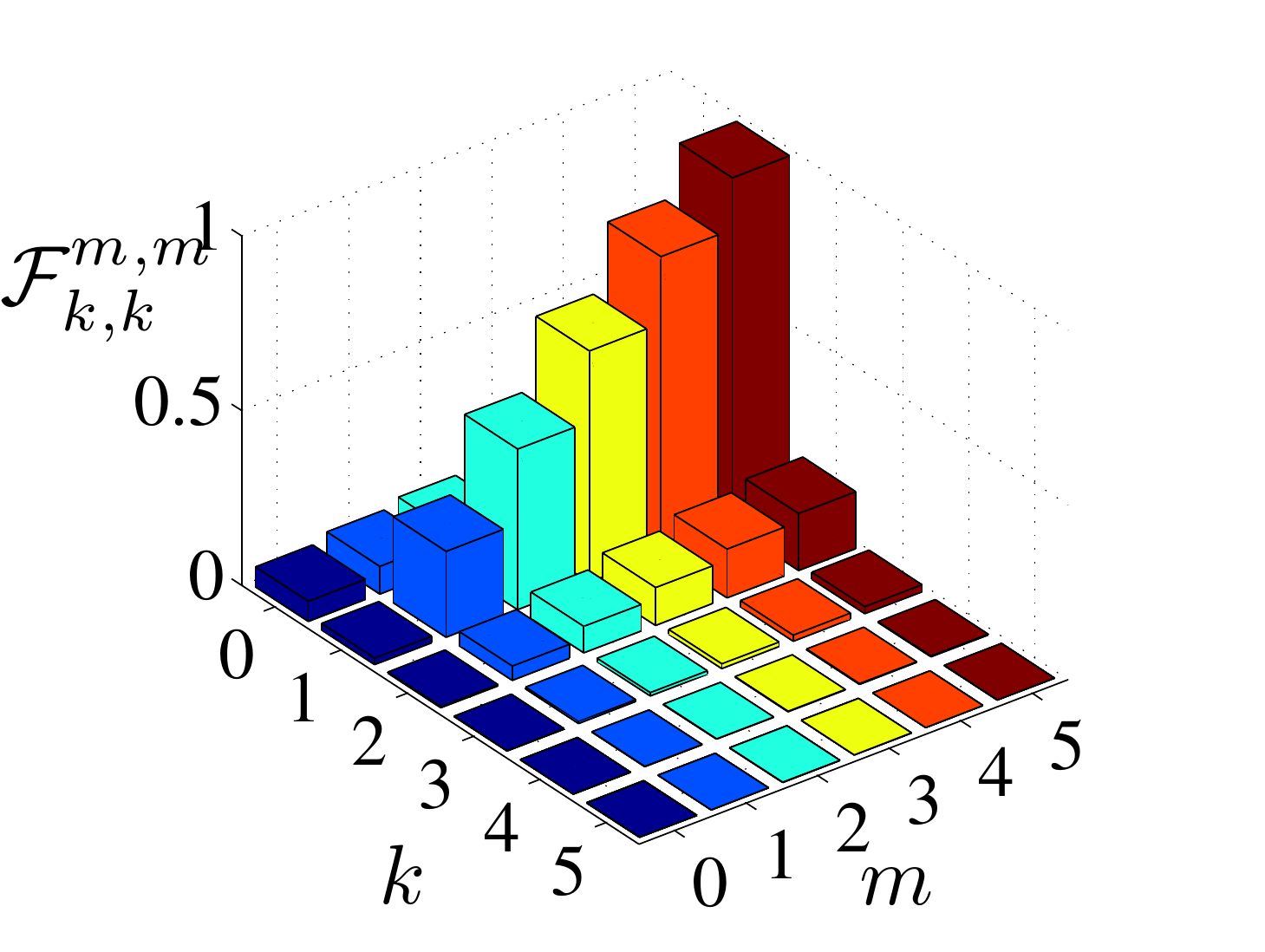}}
\subfigure[]{\includegraphics[width=0.235\textwidth]{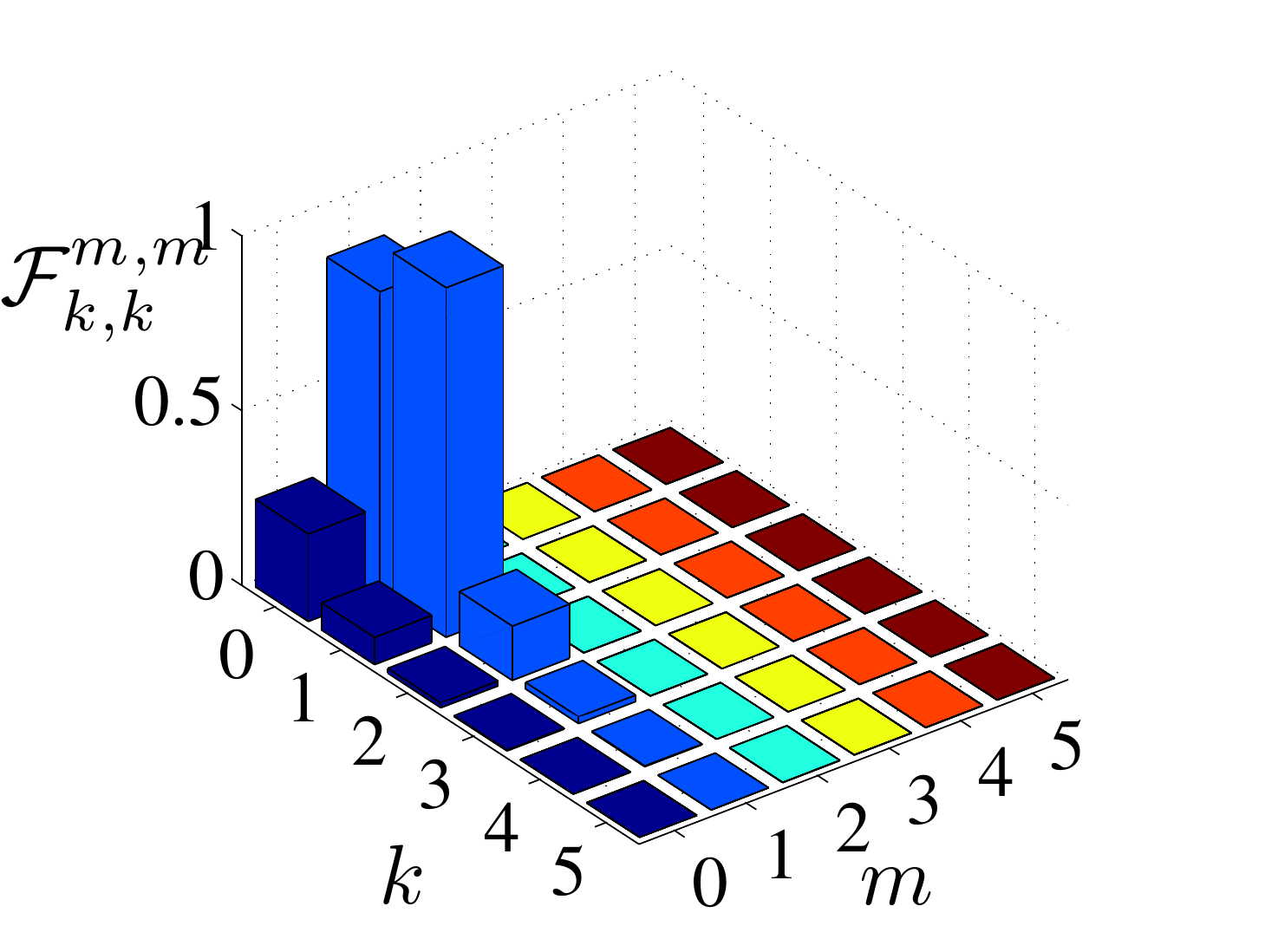}}
\subfigure[]{\includegraphics[width=0.235\textwidth]{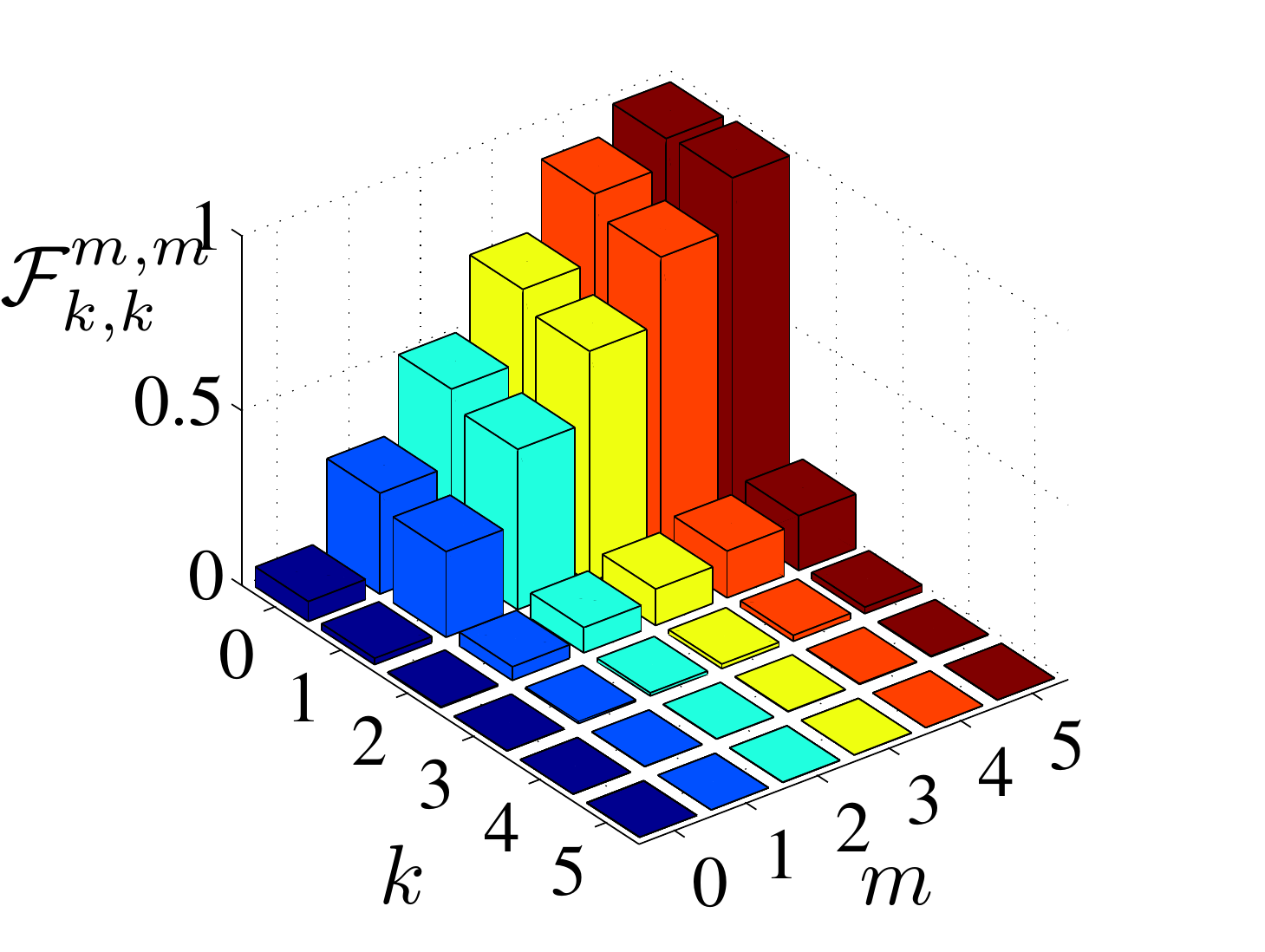}}
\caption{(a) Diagonal elements $\mathcal{F}^{m,m}_{k,k}$ of the ideal truncated noiseless amplifier process. (b) Diagonal elements $\mathcal{F}^{m,m}_{k,k}$  with non-unit detection efficiency of the APD $D_{1}$ ($\mu = 0.11)$ and lack of photon-number resolution. (c) Diagonal elements $\mathcal{F}^{m,m}_{k,k}$ with non-ideal generation of the single-photon state ($\delta = 1.089$) (d) Diagonal elements $\mathcal{F}^{m,m}_{k,k}$ including both experimental imperfections.}
\label{fig:epsilon_noiseless}
\end{figure}

Our device (Fig. \ref{fig:actual_noiseless}) is the teleportation-based amplifier proposed in Ref. \cite{Tim08}: its working principle is to use a non-maximally entangled resource -- a single photon split on an asymmetric beam splitter (A-BS) -- to perform the teleportation of a coherent state $\ket{\alpha}$. The analogue of the Bell-state measurement consists of superposing the reflected portion of the single photon with the input state on a symmetric beam-splitter (S-BS), and perform photon counting at the outputs. Successful runs are heralded by the presence of a single photon on one output and the vacuum on the other. This operation produces an output state in the form $N(\alpha)\left(\ket{0}+g\alpha\ket{1}\right)$, where $g$ is a gain factor determined by the reflexion $R$ of the A-BS, $g=\sqrt{(1-R)/R}$. For weak input intensities $\|\alpha\|\leq0.1$, this truncated expansion is a good approximation of the amplified state $\ket{g\alpha}$. Fig.\ref{fig:epsilon_noiseless} (a) shows the elements $\F^{m,m}_{k,k}$ of the corresponding map, which would normally describe the population transfer among Fock states. In this case, instead, we notice the enhancement of the single photon component and the suppression of all the higher-order terms. The complete process is a truncated form $\op{C}=g^{\op{n}}\Theta(\op{n})$, where $\Theta(\op{n})$ is 1 for $n\leq1$ and zero otherwise.

Several departures from the ideal behaviour prevent  from matching these simple predictions in the experiment, and a more refined description is then necessary. One of the main limitations is represented by single-photon detection. While the apparatus is quite robust against limited efficiency \cite{Fer10, Tim08}, it is nevertheless affected by the lack of photon-number resolution. The APD $D_{1}$, which heralds the successful events of the amplification process, will give a click each time that some light is absorbed, irrespectively on the energy. This causes triggering events which do not originate from single photon on $D_{1}$, which result in a transfer of population from higher-energy states to the one-photon Fock state, as it appears in Fig. \ref{fig:epsilon_noiseless} (b), where the corresponding tensor is shown.

\begin{figure}[b!]
\centering
\begin{minipage}{0.41\textwidth}
\subfigure[]{\includegraphics[width=1.\textwidth]{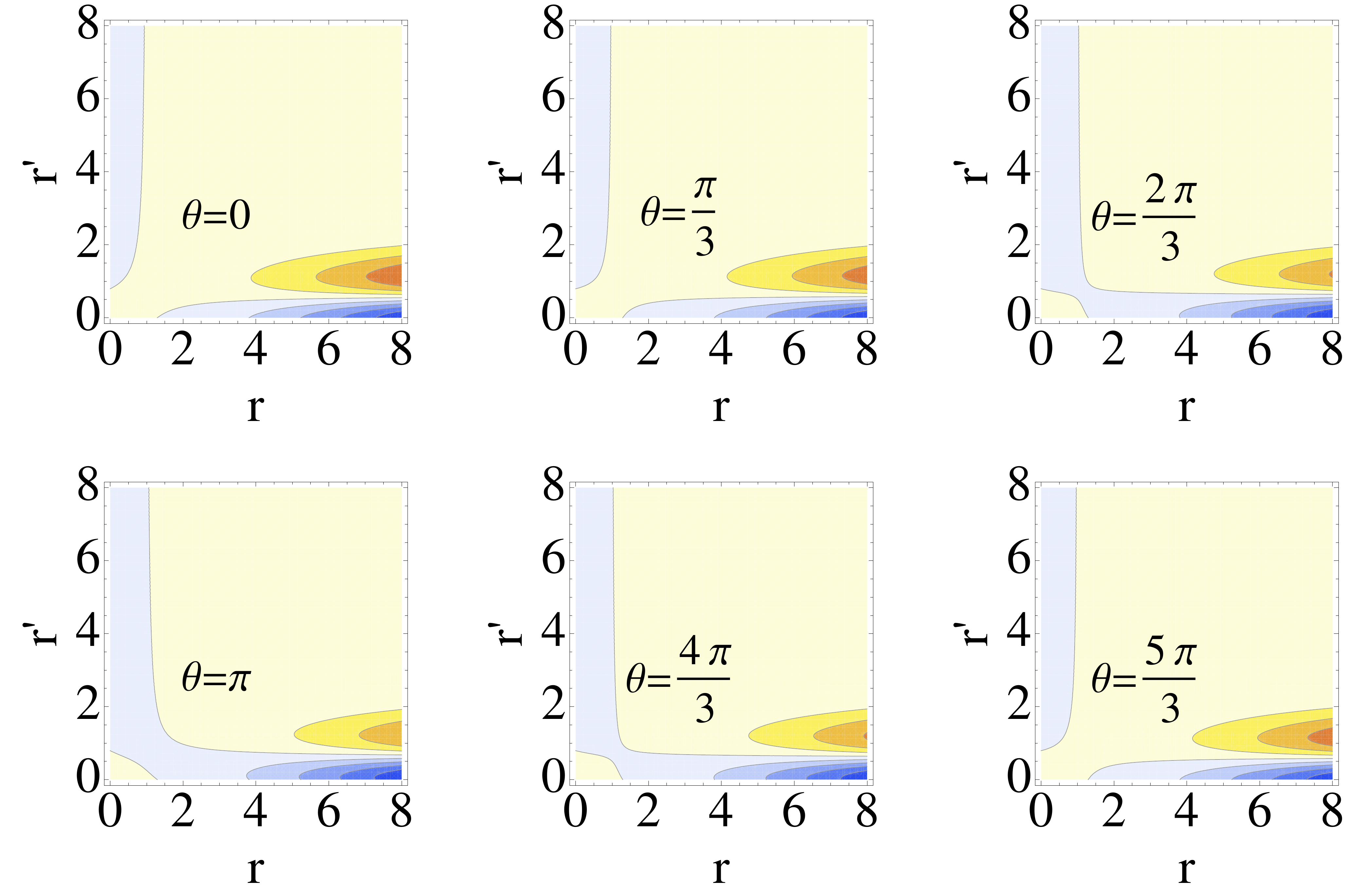}}
\subfigure[]{\includegraphics[width=1.\textwidth]{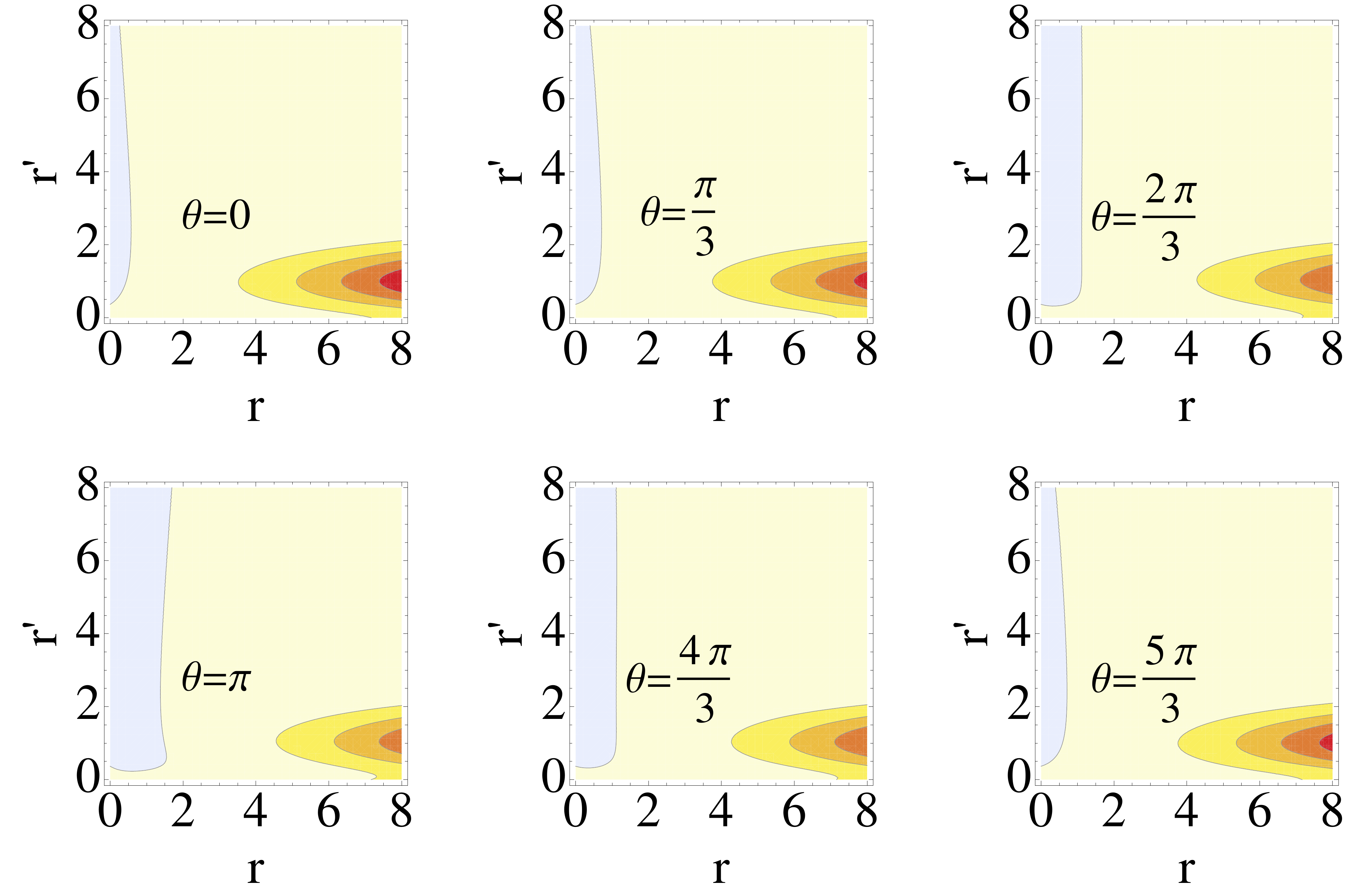}}
\end{minipage}
\begin{minipage}{0.06\textwidth}
\vspace{-10mm}
{ \includegraphics[width=1.\textwidth]{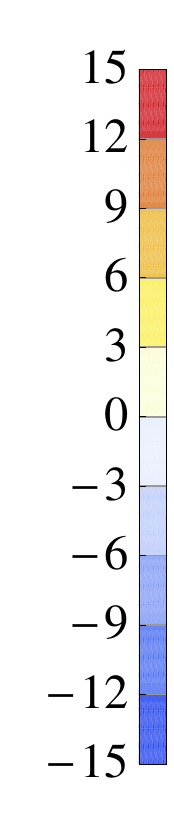} }
\vspace{0mm}
{ \includegraphics[width=1.\textwidth]{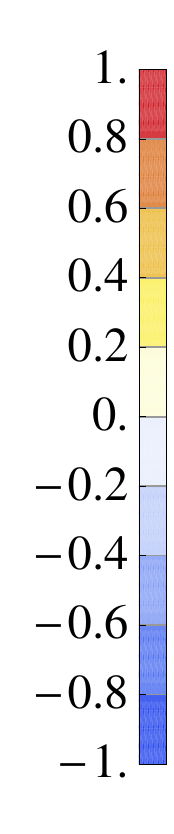} }
\vspace{-5mm}
\end{minipage}
\caption{Contour plots of $f_{\F}(r',r,\theta)$ for the noiseless amplification process as a 
function of $(r',r)$ for different values of $\theta$. (a) Low APD detection efficiency $\mu$ and
ideal generation of the single photon ($\delta = 2$). (b) Low APD detection efficiency $\mu$ and 
non-ideal generation of the single photon ($\delta = 1.089$).}
\label{fig:transfer_noiseless}
\end{figure}

Single photons are produced by down-conversion in a non-linear crystal:  whenever an avalanche photo-diode (APD) D$_0$ detects the presence of one photon, it heralds the twin photon on the correlated mode. The probabilistic nature of the emission allows for multiple-pair generation, which the APD is not able to discriminate from single-pair events.  The output state will not be a pure single photon, but will present contributions from higher order terms. Furthermore, one needs to consider that the matching of the pump field with the observed modes will not be perfect. This results in excess noise in both the conditioning and signal modes, spoiling even more  the quality of our single photon state.  This can be assessed in the experiment by measuring the quadrature distributions \cite{Our06}, and is described by a parameter $\delta$ \cite{Our06a,Fer10a}, ranging from $\delta = 2$ for a pure heralded single-photon state to $\delta = 0$ for a thermal state. The effect on the process is illustrated in Fig.\ref{fig:epsilon_noiseless} (c) showing how higher-order number from multiple-pair emission can end up being populated.

Considering both imperfection sources provides an exhaustive model of our experiment: its accuracy can be checked by calculating the fidelities between the experimental density matrices \cite{Fer10} and the prediction of the model (right inset of Fig.\ref{fig:actual_noiseless}), showing an average figure of $\sim 99.5\%$, for weak intensities $\|\alpha\|\leq1$. The results are summarised in Fig. \ref{fig:epsilon_noiseless}(d): the two mechanisms take place independently and cause a lesser gain than expected, and the presence of noise in the amplified states. 

The transfer function can be decomposed in a correctly heralded  and a faulty one.  For the correctly heralded transfer function, one could start with the beamsplitters: the asymmetric one is obtained by composing, with the use of Eq. \ref{eq:comb-map}, the beamsplitter transfer function with the Wigner function of the experimental single photon and the vacuum. On the same way, the transfer function for the part containing the symmetric beamsplitter is determined by composing the beamsplitter transfer function with the transfer function of the APD (Wigner function of the projection operator multiplied by $2\pi$ composed with an attenuation transfer function) on one output and an attenuation (modelising the mode-matching on one input), and then tracing on the remaining output mode. The final correctly heralded transfer function is obtained by composing those two transfer function. The faulty transfer function is simply determined by composing an attenuation transfer function (taking into account the mode-matching and the symmetric beamsplitter) with the APD transfer function.

While an inspection in the Fock basis can be informative, it does not lead to the most natural description of a process for continuous-variable states; also, from a practical point, it might be cumbersome to verify some properties such as the Gaussianity of the process, or its nonclassicality from the expression of the $\F$ tensor. On this purpose, a useful approach consists in inspecting the trend of the associated transfer function: while this object generally acts on pairs of two-dimensional vectors $\vec{r}=(x,p)$ and $\vec{r}'=(x',p')$, for phase-invariant processes -- as it is the case for the noiseless amplifier -- the transfer function can only depend on $r=\sqrt{x^{2}+p^{2}}$, $r'=\sqrt{x^{'2}+p^{'2}}$, and $\theta = \cos^{-1}\left(\frac{\vec{r} \cdot \vec{r}'}{r r'}\right)$. The transfer function for the noiseless amplifier is presented in Fig.\ref{fig:transfer_noiseless}, comparing the cases when an ideal single photon is used as ancilla and with the actual resources. Non-classical features are clear in the ideal limit, in which negative values appears around $r'{=}0$. However, in experimental conditions, these signatures are smoothed by the imperfections of the set-up, though there remains negative region. In more details, we observe how low values of the transfer function for high $r'$ correspond to the saturation of the amplifier, i.e. the impossibility of having more than one photon at the output. The negative peak determines the non-Gaussianity of the output states by causing a small negative region in the Wigner function of the output state; nevertheless, this feature vanishes rather quickly with the imperfection and is not visible in realistic output state, as it appears in Fig.\ref{fig:transfer_noiseless}. 
We can notice that the region for small $r$ and $r'$ is quite different around $\theta=0$ and $\theta=\pi$, this is due to the fact that the amplifier keeps the phase of the "small" states, whereas for bigger value of $r$ and $r'$ the transfer function is almost independent of $\theta$ since the higher photon-number terms in the input state often trigger the heralding leading to a single photon (with losses) in the output state.
A last remark concerns the increasing peak with $r$ and the different scales between the two maps originating from the variation of the success probability.

\section{Example 2: photon addition}

The extreme negative value of a Wigner function can be used as a quantifier of its non-classicality \cite{Bar10}. However, an intuitive extension of such a reasoning to the transfer functions would be severely affected by the probabilistic character of the process itself. Here we illustrate these considerations in a second example: the single-photon addition ($\op{C}{=}\op{a}^\dag$). As above, our description is mediated by a model of the physical process.

\begin{figure}[t]
\centering
\includegraphics[width=0.5\textwidth]{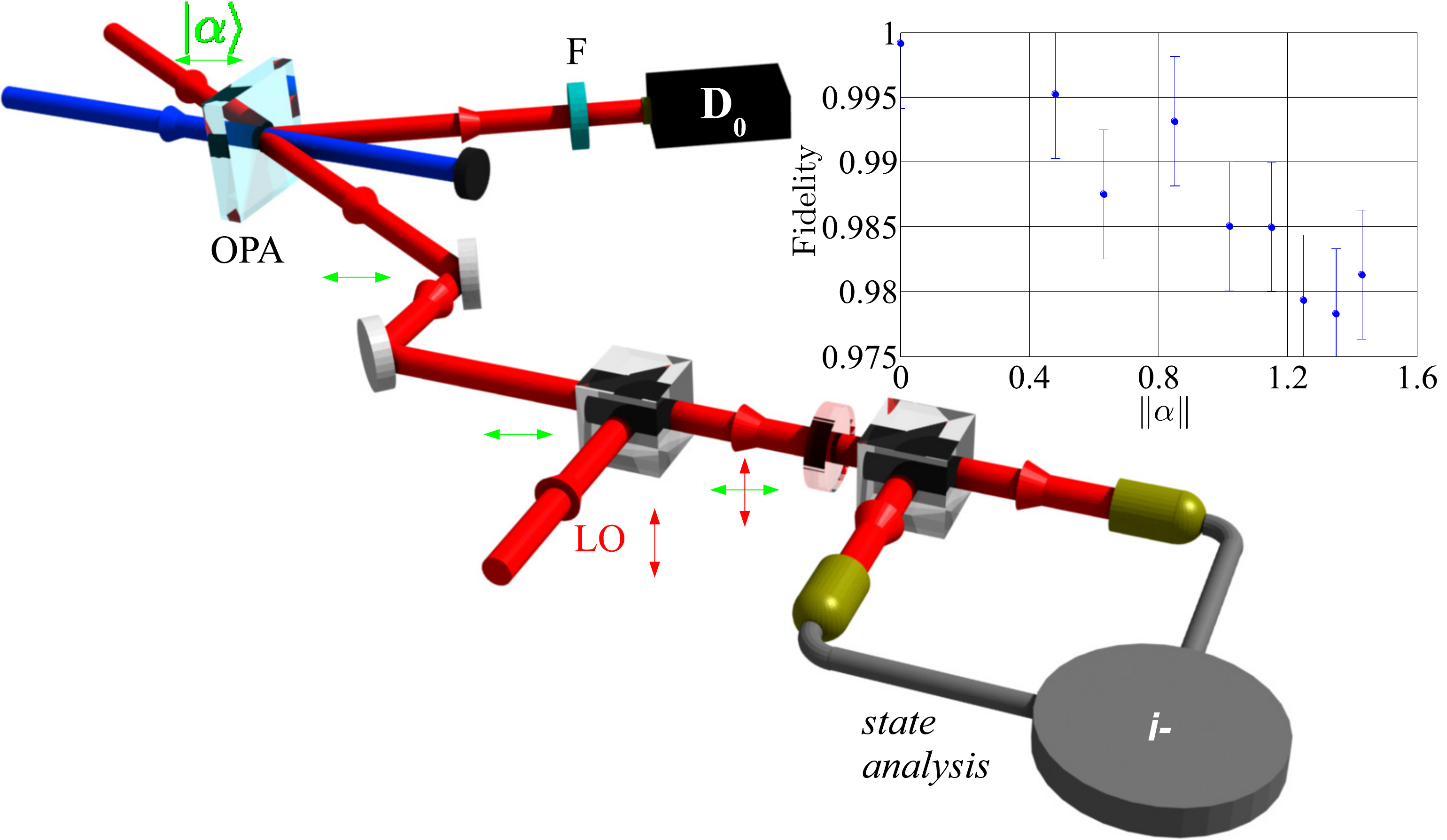}
\caption{Layout of the single-photon addition experiment  \cite{Bar10}. A single photon
is conditionally added upon detection of a single photon on detector $D_{0}$.  As above, the beam-splitters exploit
polarization (double sided arrows in the figure).Inset: fidelities between the experimental density matrices \cite{Bar10} 
and the prediction of the model.}
\label{fig:actual_photonaddition}
\end{figure}

In our implementation, photon addition is achieved by feeding the input state in an optical parametric amplifier (OPA) driven at low gain $g{=}\cosh^2 \chi$, where $\chi$ measures the non-linear interaction strength and it is proportional to the pump intensity. To the first order, this process adds a photon pair shared by the signal mode, and a correlated mode, on which an APD $D_0$ is placed; due to the nondeterministc nature of the process, the successful events are triggered by a detection event from $D_0$ (Fig.~\ref{fig:actual_photonaddition}). This method has been introduced in \cite{Zav04}, and then adopted for tests of the commutation rules \cite{Par07,Zav09}, and the analysis of non-classicality \cite{Zav07} and non-Gaussianity \cite{Bar10}.

\begin{figure}[b]
\centering
\subfigure[]{\includegraphics[width=0.235\textwidth]{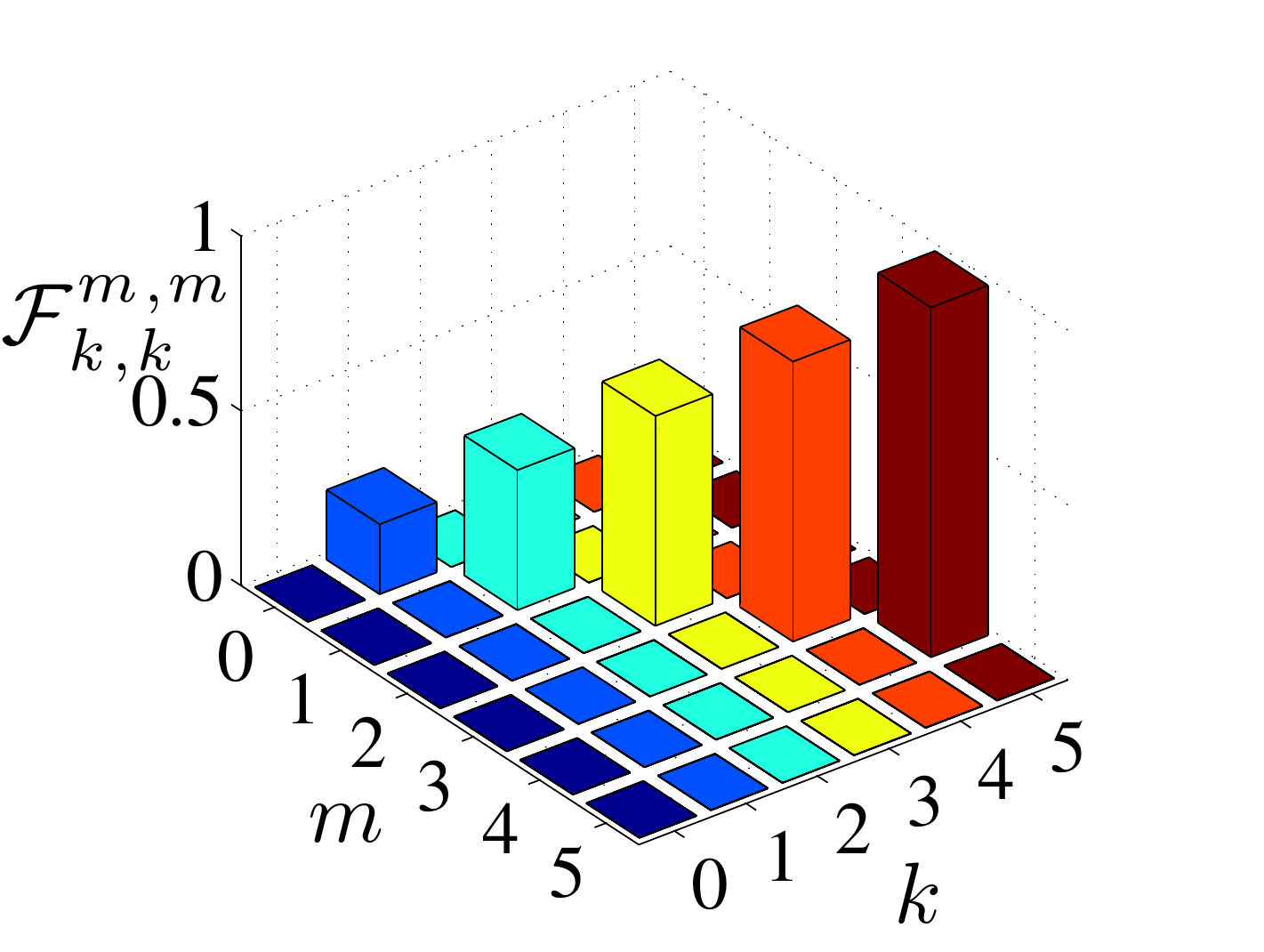}}
\subfigure[]{\includegraphics[width=0.235\textwidth]{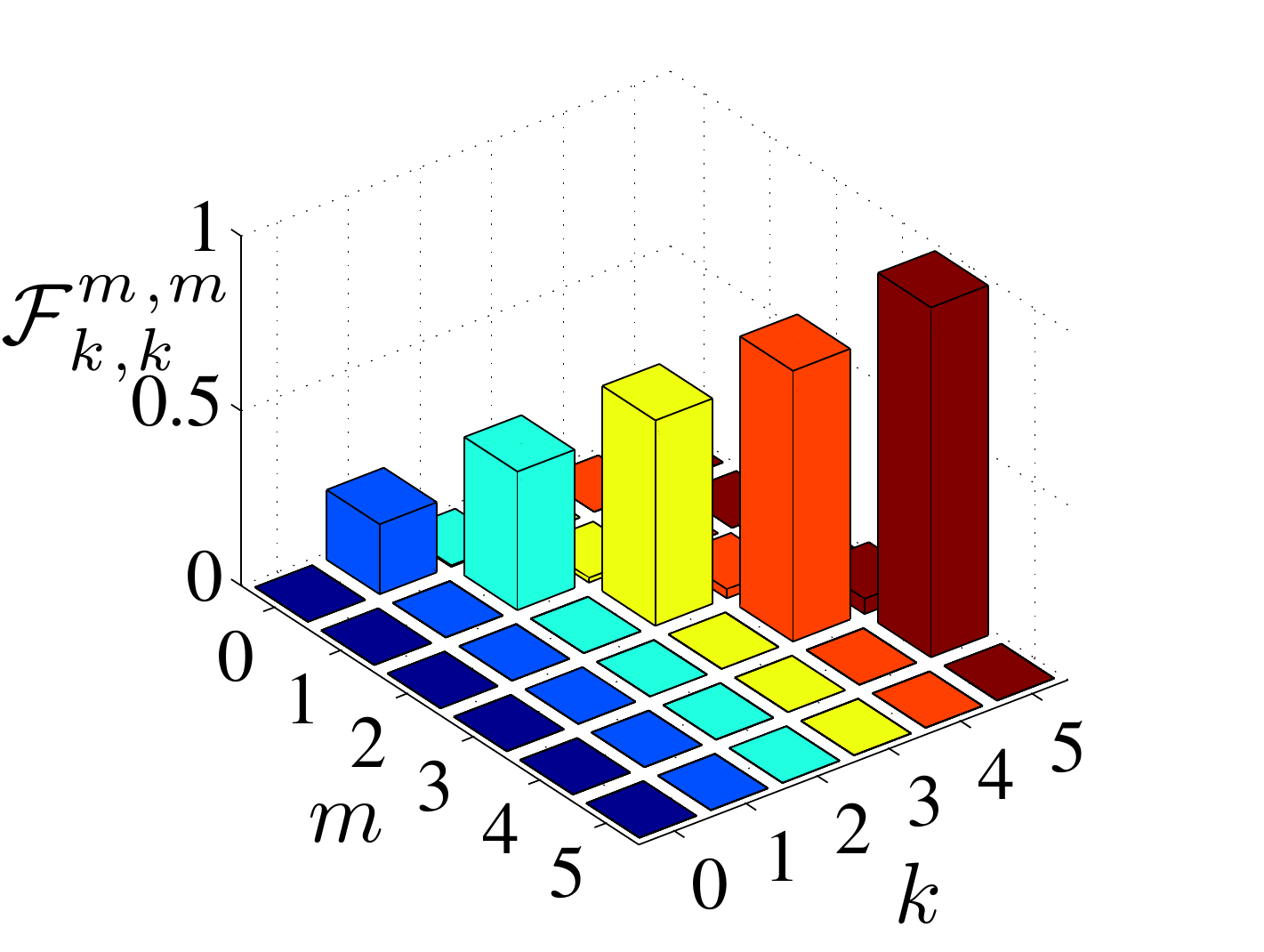}}
\subfigure[]{\includegraphics[width=0.235\textwidth]{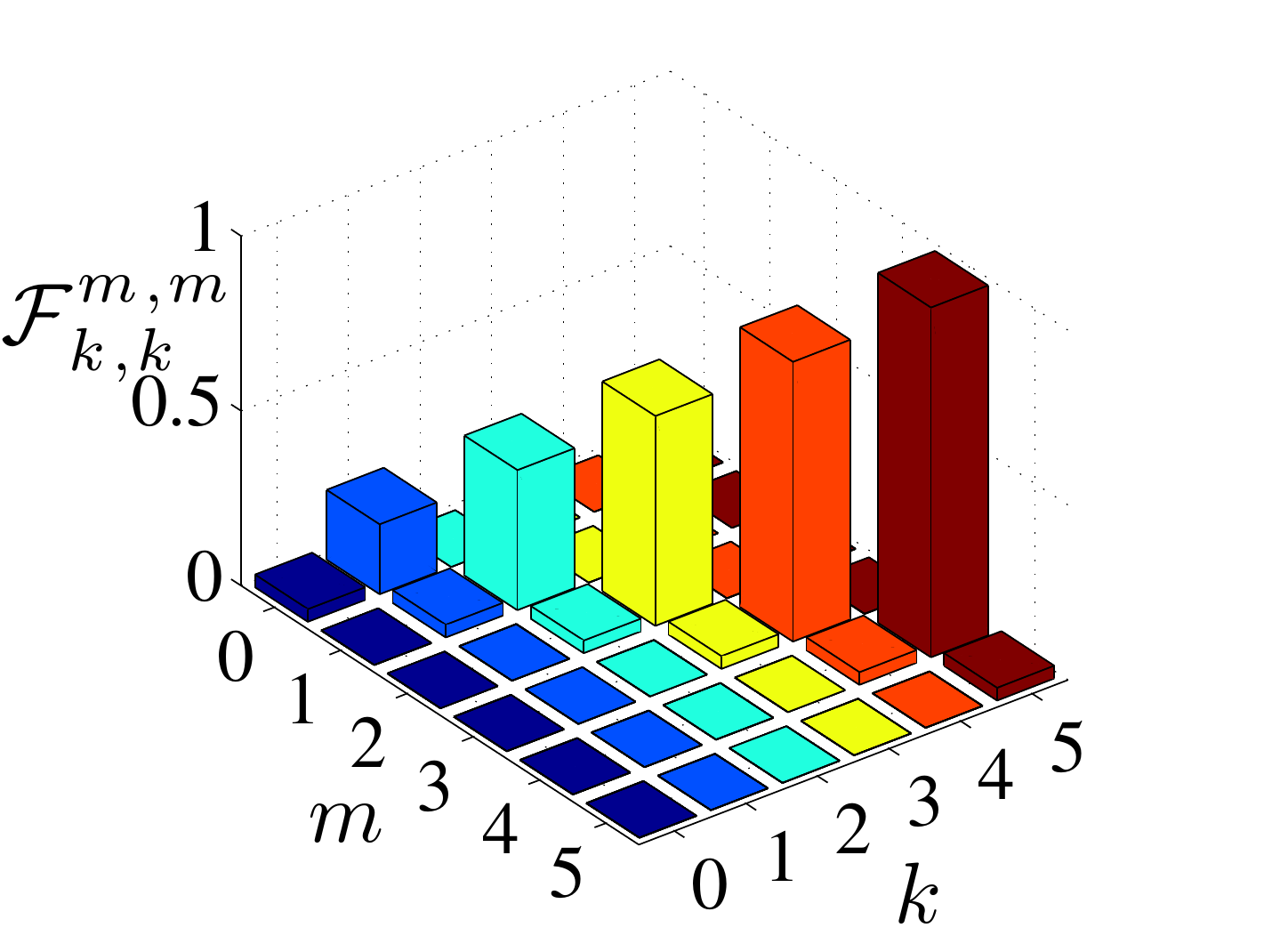}}
\subfigure[]{\includegraphics[width=0.235\textwidth]{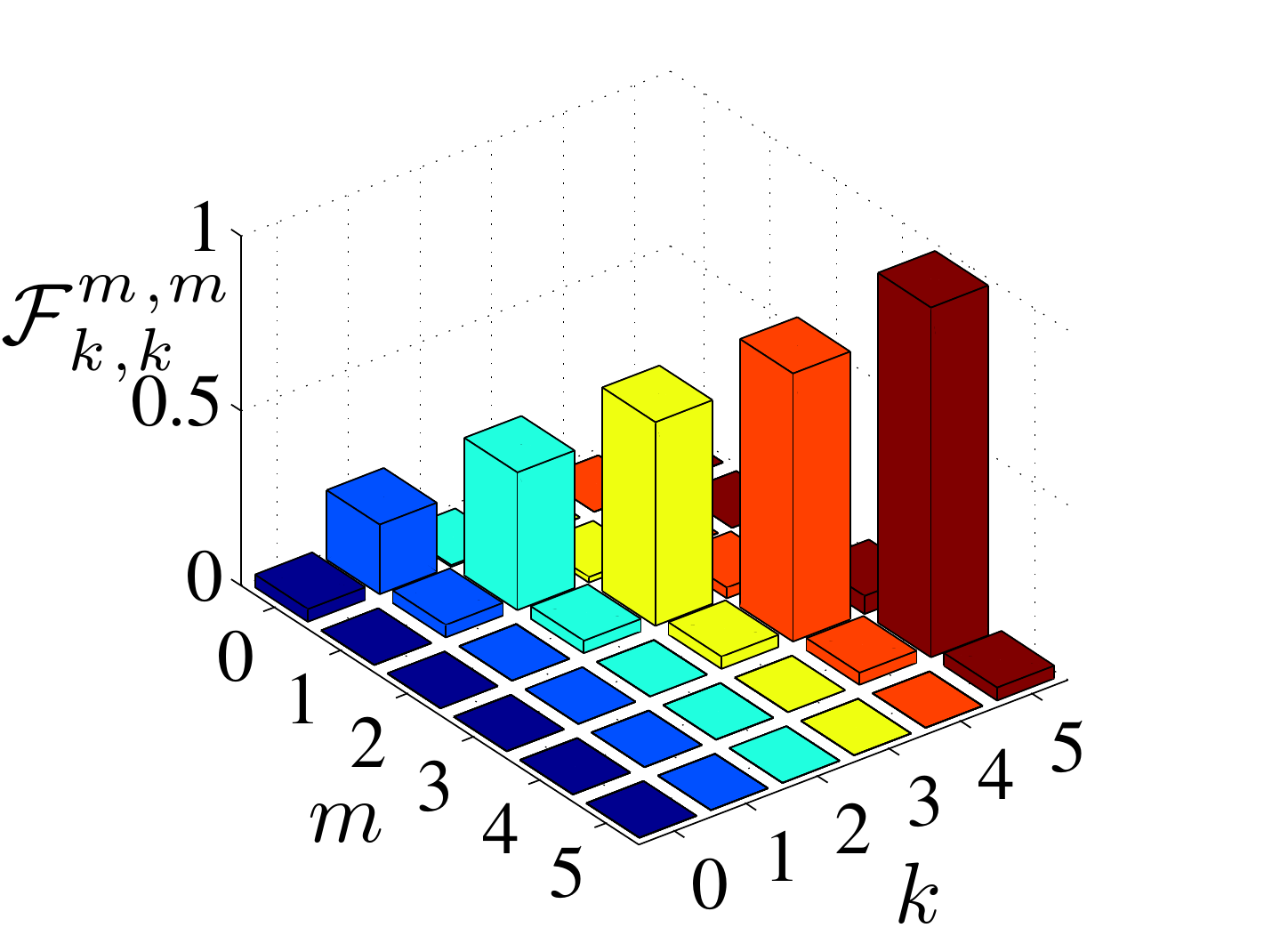}}
\caption{(a) Diagonal elements $\mathcal{F}^{m,m}_{k,k}$ of the ideal photon addition process. (b) Diagonal elements $\mathcal{F}^{m,m}_{k,k}$  for the case of a conditioned OPA driven at $\chi{=}0.105$. (c) Diagonal elements $\mathcal{F}^{m,m}_{k,k}$ with a parasitic gain $\gamma{=}0.425$ and very low gain (d) Diagonal elements $\mathcal{F}^{m,m}_{k,k}$ including both experimental imperfections.}
\label{fig:epsilon_addition}
\end{figure}

The main source of noise here can be identified in the imperfect matching between the pump and the signal modes, which results in a parasite gain $h{=}\cosh^2\gamma \chi$. An estimate for these two gains, taken from a fit of their non-Gaussianity \cite{Bar10}, is $\chi{=}0.105$, and $\gamma{=}0.425$ \cite{Bar10}. A third imperfection arises from the fact that spurious events might happen at $D_0$, due either to dark counts or clicks originating from non-matching modes. 
The average fidelity between modelled and the reconstructed states using coherent states as inputs is satisfactory, although the data might be affected by some extra noise likely due to low-frequency fluctuations of the average level of the homodyne current.

As it appears from the comparison of Figs.~\ref{fig:epsilon_addition}a and b, the gain $\chi$ is chosen to be sufficiently low so that two-pair events are not significant: the transfer of population by more than one photon is low. On the other hand, the effect of the parasite gain seems as important: the sheer effect is the presence of uncorrelated clicks at $D_0$ that leave the state unchanged. This corresponds to the diagonal terms in Fig.~\ref{fig:epsilon_addition}c, considered in the limit of extremely low gain $\chi\rightarrow0$. The overall process simply results in the presence of these two imperfections. For the sake of simplicity we have only considered the case of low detection efficiency at $D_0$ (Fig.~\ref{fig:epsilon_addition}d).  In this example, the adoption of the quantum map formalism reveals to be particularly clear and useful for the analysis of the process: not only it confirms our intuition about the behaviour of parasite processes, but also give us a way of quantifying their effect in a way that does not depend on the particular input.

\begin{figure}[t]
\centering
\subfigure[]{\includegraphics[width=0.235\textwidth]{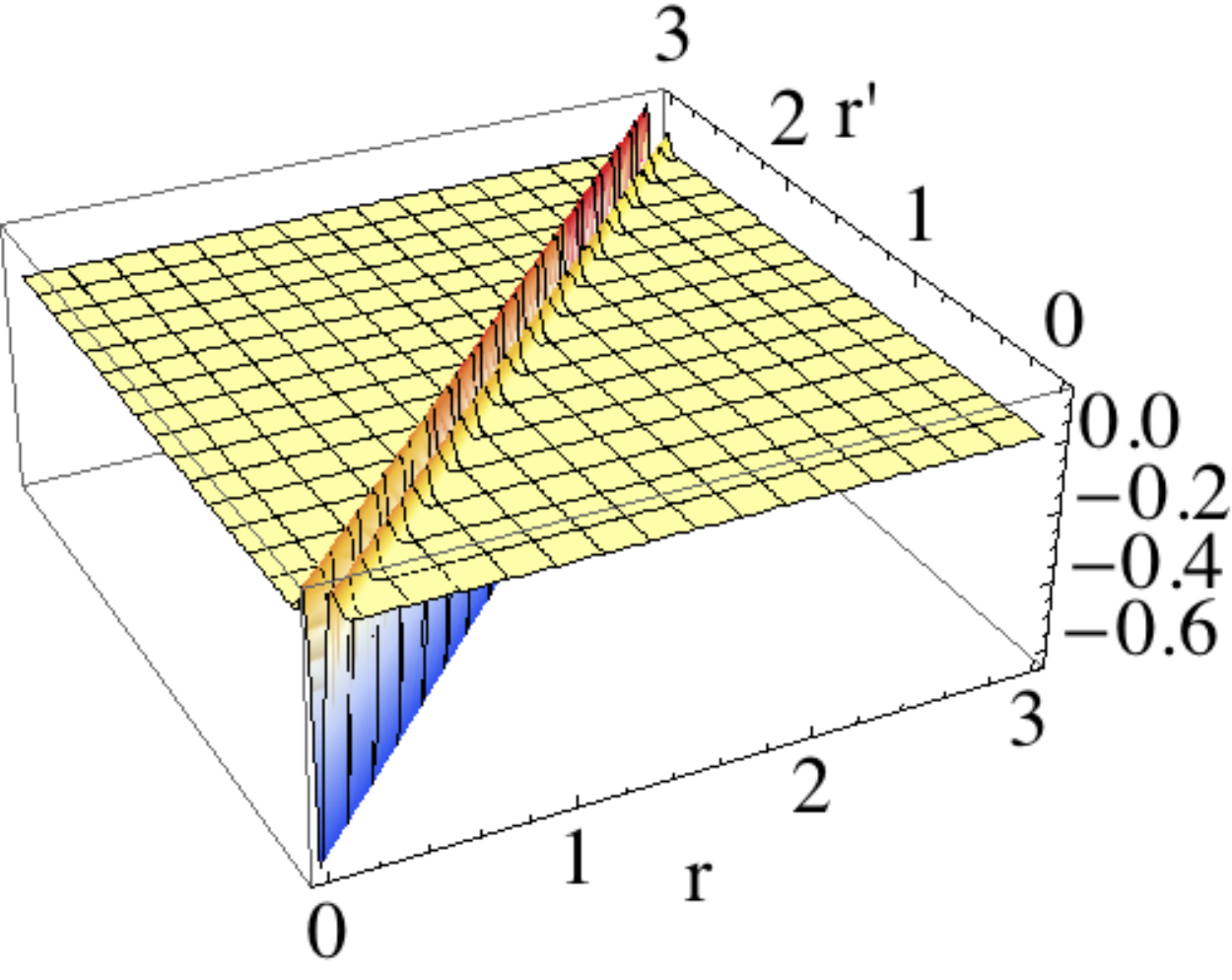}}
\subfigure[]{\includegraphics[width=0.235\textwidth]{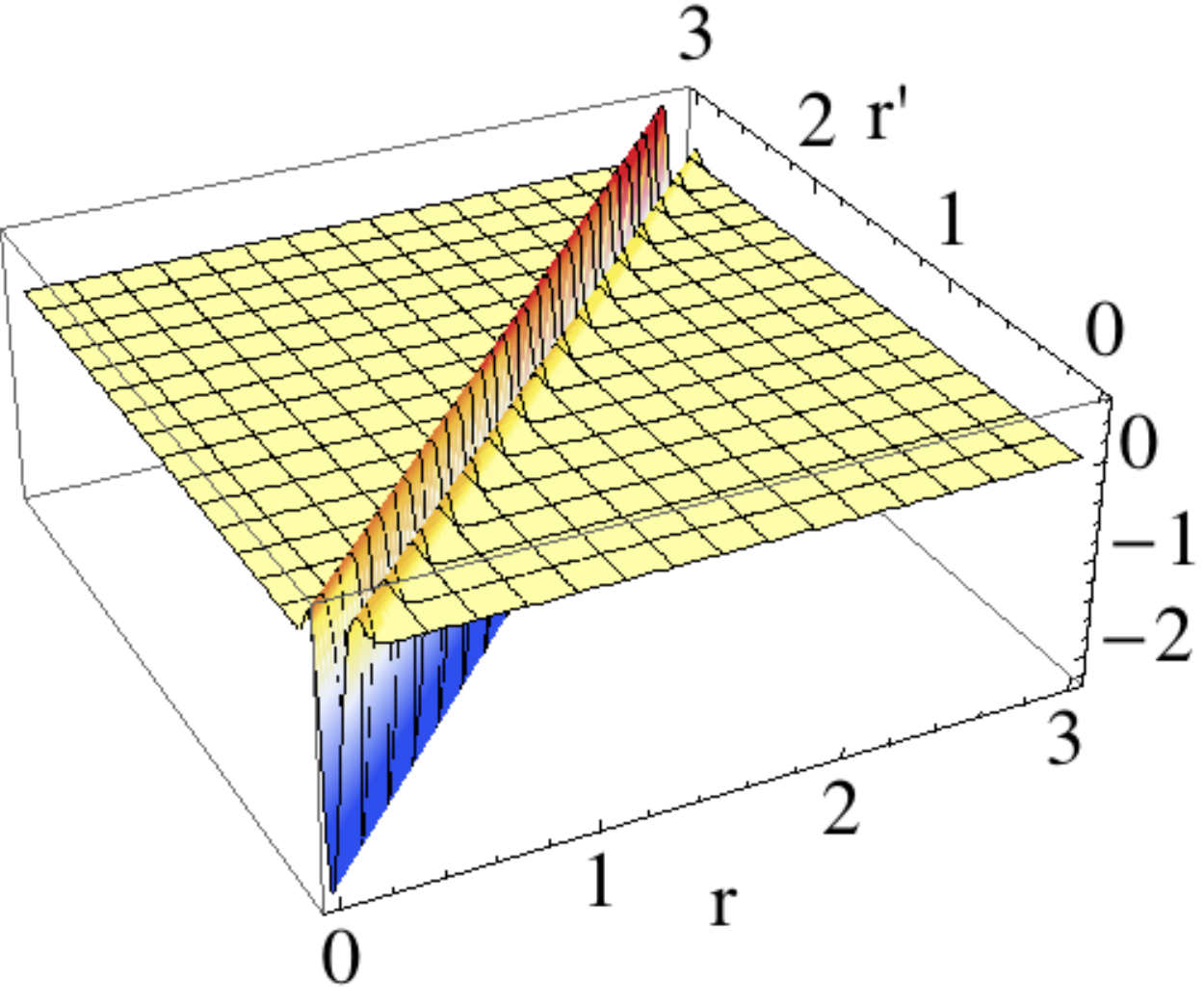}}
\caption{Plots of $f_{\F}(r',r,\theta)$ for the photon addition process as a 
function of $(r',r)$ for $\theta=0$. (a) With ideal OPA (driven at $\chi{=}0.105$) and photon counter. (b) With a parasitic gain $\gamma{=}0.425$ and APD.}
\label{fig:transfer_photon}
\end{figure}.

\begin{figure}[t]
\centering
\subfigure[]{\includegraphics[width=0.235\textwidth]{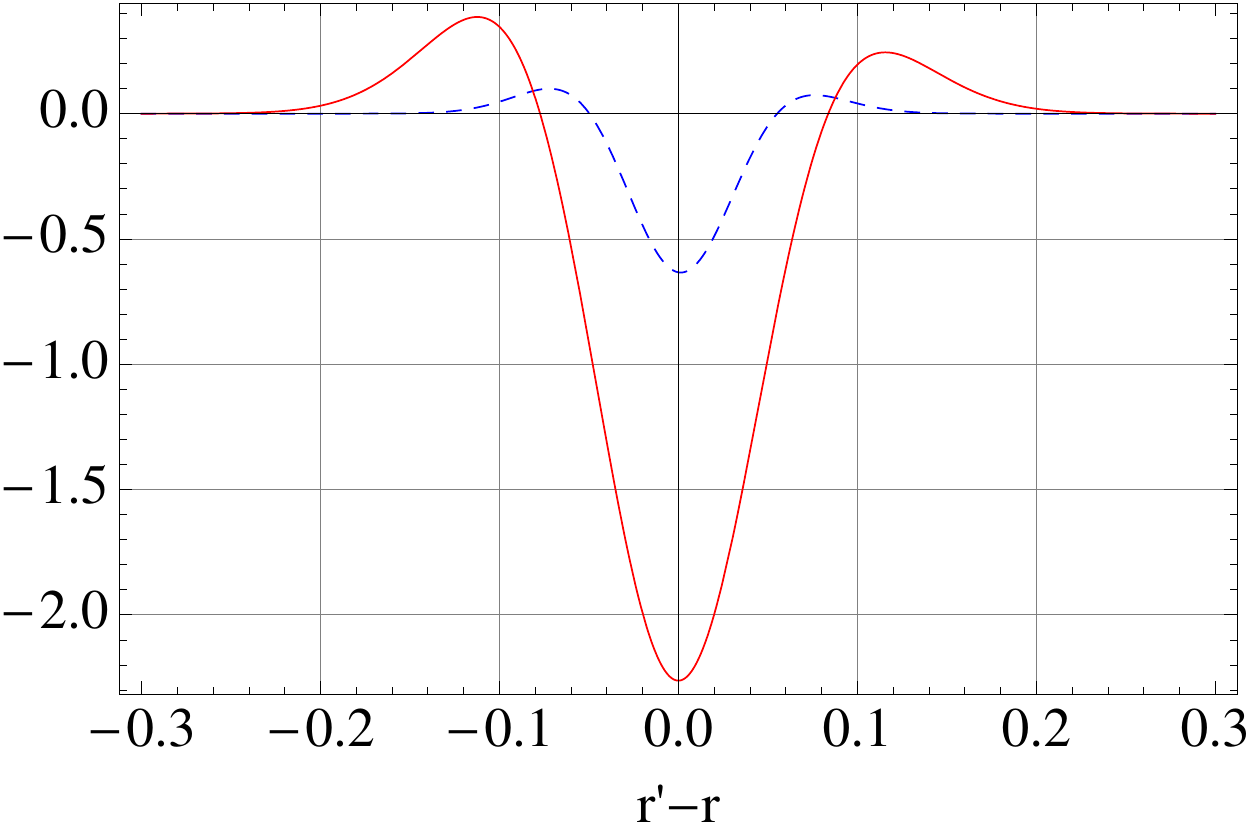}}
\subfigure[]{\includegraphics[width=0.235\textwidth]{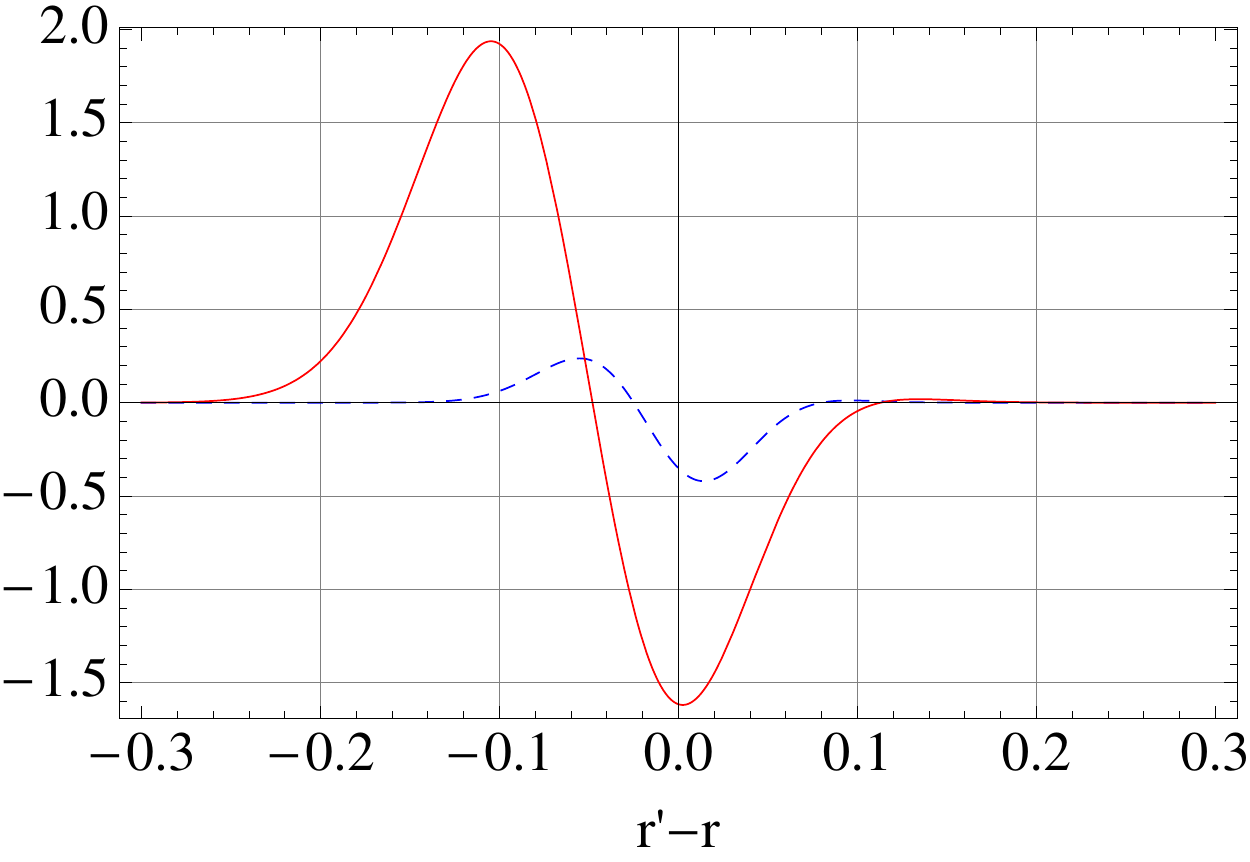}}
\caption{Plots of $f_{\F}(r',r,\theta)$ for the photon addition process as a 
function of $r'-r$. The plain line correspond to the experimental conditions ( $\gamma{=}0.425$ and APD), and the dashed line to the ideal case. (a) for $\theta=0$ and $r+r'=2$ (b)  for $\theta=0$ and $r+r'=20$}
\label{fig:transfer_photon_profil}
\end{figure}

The correctly heralded transfer function is easy to determine: it is only a composition of the parametric downconversion transfer function with the Wigner function of vacuum on one input and the APD transfer function and attenuation transfer function on the outputs. 
As for the preceding example, we can use the radial symmetry of the process to express the transfer function in the simplest form $f_{\F}(r',r,\theta)$. In Fig.\ref{fig:transfer_photon} we show the transfer function $f_{\F}(r',r,0)$ as a function of $(r',r)$, as this contains most information about the physics: there we compare the case of low gain $r$ and photon-number discrimination, and the full model of our experiment. For both cases, the transfer function has non-zero values only around $r'=r$ and for $\theta=0$: this indicates that the amplitude and phase are mostly unchanged by the process. The increase of the positive peak with the amplitude corresponds to the growth of the success rate with the number of photons, and is more visible in the second graph because of the inability to discriminate the photon number. Finally, the negative peak introduce a negative part in the resulting Wigner function and is a sign of the non-gaussianity of the process. We can note that the difference between the two scales (Fig. \ref{fig:transfer_photon_profil}) are only due to the differences in the success rate, even for the negative peak. It implies that the size of the negative peak can not be readily used to quantify the quantumness of a map.

\section{Conclusion}

We have inspected two important processes for continuous-variable states with the formalism of quantum maps: these can convey interesting physical information about the process independently on the state.  The adoption of a description in terms of transfer functions offers a compact, insightful view of the process, along the same lines of what happens with the Wigner function for quantum states. We have applied this method to the description of the realistic operation of a noiseless amplifier, and of a  photon-adder, evidenciating how experimental imperfections shape the features of the transfer function.

We thank Ph. Grangier, M.G. Genoni and M.G.A. Paris for discussion. We acknowledge support from the EU project ANR ERA-Net CHISTERA HIPERCOM. MB is supported by the Marie Curie contract PIEF-GA-2009-236345- PROMETEO.

\bibliography{bibliography_QPT}

\begin{thebibliography}{32}%
\makeatletter
\providecommand \@ifxundefined [1]{%
 \@ifx{#1\undefined}
}%
\providecommand \@ifnum [1]{%
 \ifnum #1\expandafter \@firstoftwo
 \else \expandafter \@secondoftwo
 \fi
}%
\providecommand \@ifx [1]{%
 \ifx #1\expandafter \@firstoftwo
 \else \expandafter \@secondoftwo
 \fi
}%
\providecommand \natexlab [1]{#1}%
\providecommand \enquote  [1]{``#1''}%
\providecommand \bibnamefont  [1]{#1}%
\providecommand \bibfnamefont [1]{#1}%
\providecommand \citenamefont [1]{#1}%
\providecommand \href@noop [0]{\@secondoftwo}%
\providecommand \href [0]{\begingroup \@sanitize@url \@href}%
\providecommand \@href[1]{\@@startlink{#1}\@@href}%
\providecommand \@@href[1]{\endgroup#1\@@endlink}%
\providecommand \@sanitize@url [0]{\catcode `\\12\catcode `\$12\catcode
  `\&12\catcode `\#12\catcode `\^12\catcode `\_12\catcode `\%12\relax}%
\providecommand \@@startlink[1]{}%
\providecommand \@@endlink[0]{}%
\providecommand \url  [0]{\begingroup\@sanitize@url \@url }%
\providecommand \@url [1]{\endgroup\@href {#1}{\urlprefix }}%
\providecommand \urlprefix  [0]{URL }%
\providecommand \Eprint [0]{\href }%
\providecommand \doibase [0]{http://dx.doi.org/}%
\providecommand \selectlanguage [0]{\@gobble}%
\providecommand \bibinfo  [0]{\@secondoftwo}%
\providecommand \bibfield  [0]{\@secondoftwo}%
\providecommand \translation [1]{[#1]}%
\providecommand \BibitemOpen [0]{}%
\providecommand \bibitemStop [0]{}%
\providecommand \bibitemNoStop [0]{.\EOS\space}%
\providecommand \EOS [0]{\spacefactor3000\relax}%
\providecommand \BibitemShut  [1]{\csname bibitem#1\endcsname}%
\let\auto@bib@innerbib\@empty
\bibitem [{\citenamefont {Leonhardt}(1998)}]{Leo97}%
  \BibitemOpen
  \bibfield  {author} {\bibinfo {author} {\bibfnamefont {U.}~\bibnamefont
  {Leonhardt}},\ }\href@noop {} {\emph {\bibinfo {title} {Measuring the quantum
  state of light}}}\ (\bibinfo  {publisher} {Cambridge University Press},\
  \bibinfo {year} {1998})\BibitemShut {NoStop}%
\bibitem [{\citenamefont {James}\ \emph {et~al.}(2001)\citenamefont {James},
  \citenamefont {White}, \citenamefont {Kwiat},\ and\ \citenamefont
  {Munro}}]{Jam01}%
  \BibitemOpen
  \bibfield  {author} {\bibinfo {author} {\bibfnamefont {D.~F.~V.}\
  \bibnamefont {James}}, \bibinfo {author} {\bibfnamefont {A.~G.}\ \bibnamefont
  {White}}, \bibinfo {author} {\bibfnamefont {P.~G.}\ \bibnamefont {Kwiat}}, \
  and\ \bibinfo {author} {\bibfnamefont {W.~J.}\ \bibnamefont {Munro}},\
  }\href@noop {} {\bibfield  {journal} {\bibinfo  {journal} {Phys. Rev. A}\
  }\textbf {\bibinfo {volume} {64}},\ \bibinfo {pages} {052312} (\bibinfo
  {year} {2001})}\BibitemShut {NoStop}%
\bibitem [{\citenamefont {O'Brien}\ \emph {et~al.}(2004)\citenamefont
  {O'Brien}, \citenamefont {Pryde}, \citenamefont {Gilchrist}, \citenamefont
  {James}, \citenamefont {Langford}, \citenamefont {Ralph},\ and\ \citenamefont
  {White}}]{OBr04}%
  \BibitemOpen
  \bibfield  {author} {\bibinfo {author} {\bibfnamefont {J.~L.}\ \bibnamefont
  {O'Brien}}, \bibinfo {author} {\bibfnamefont {G.~J.}\ \bibnamefont {Pryde}},
  \bibinfo {author} {\bibfnamefont {A.}~\bibnamefont {Gilchrist}}, \bibinfo
  {author} {\bibfnamefont {D.~F.~V.}\ \bibnamefont {James}}, \bibinfo {author}
  {\bibfnamefont {N.~K.}\ \bibnamefont {Langford}}, \bibinfo {author}
  {\bibfnamefont {T.~C.}\ \bibnamefont {Ralph}}, \ and\ \bibinfo {author}
  {\bibfnamefont {A.~G.}\ \bibnamefont {White}},\ }\href@noop {} {\bibfield
  {journal} {\bibinfo  {journal} {Phys. Rev. Lett.}\ }\textbf {\bibinfo
  {volume} {93}},\ \bibinfo {pages} {080502} (\bibinfo {year}
  {2004})}\BibitemShut {NoStop}%
\bibitem [{\citenamefont {Lobino}\ \emph {et~al.}(2008)\citenamefont {Lobino},
  \citenamefont {Korystov}, \citenamefont {Kupchak}, \citenamefont {Figueroa},
  \citenamefont {Sanders},\ and\ \citenamefont {Lvovsky}}]{Lob08}%
  \BibitemOpen
  \bibfield  {author} {\bibinfo {author} {\bibfnamefont {M.}~\bibnamefont
  {Lobino}}, \bibinfo {author} {\bibfnamefont {D.}~\bibnamefont {Korystov}},
  \bibinfo {author} {\bibfnamefont {C.}~\bibnamefont {Kupchak}}, \bibinfo
  {author} {\bibfnamefont {E.}~\bibnamefont {Figueroa}}, \bibinfo {author}
  {\bibfnamefont {B.~C.}\ \bibnamefont {Sanders}}, \ and\ \bibinfo {author}
  {\bibfnamefont {A.~I.}\ \bibnamefont {Lvovsky}},\ }\href@noop {} {\bibfield
  {journal} {\bibinfo  {journal} {Science}\ }\textbf {\bibinfo {volume}
  {322}},\ \bibinfo {pages} {563} (\bibinfo {year} {2008})}\BibitemShut
  {NoStop}%
\bibitem [{\citenamefont {Lundeen}\ \emph {et~al.}(2008)\citenamefont
  {Lundeen}, \citenamefont {Feito}, \citenamefont {Coldenstrodt-Ronge},
  \citenamefont {Pregnell}, \citenamefont {Silberhorn}, \citenamefont {Ralph},
  \citenamefont {Eisert}, \citenamefont {Plenio},\ and\ \citenamefont
  {Walmsley}}]{Lun08}%
  \BibitemOpen
  \bibfield  {author} {\bibinfo {author} {\bibfnamefont {J.~S.}\ \bibnamefont
  {Lundeen}}, \bibinfo {author} {\bibfnamefont {A.}~\bibnamefont {Feito}},
  \bibinfo {author} {\bibfnamefont {H.}~\bibnamefont {Coldenstrodt-Ronge}},
  \bibinfo {author} {\bibfnamefont {K.~L.}\ \bibnamefont {Pregnell}}, \bibinfo
  {author} {\bibfnamefont {C.}~\bibnamefont {Silberhorn}}, \bibinfo {author}
  {\bibfnamefont {T.~C.}\ \bibnamefont {Ralph}}, \bibinfo {author}
  {\bibfnamefont {J.}~\bibnamefont {Eisert}}, \bibinfo {author} {\bibfnamefont
  {M.~B.}\ \bibnamefont {Plenio}}, \ and\ \bibinfo {author} {\bibfnamefont
  {I.~A.}\ \bibnamefont {Walmsley}},\ }\href@noop {} {\bibfield  {journal}
  {\bibinfo  {journal} {Nature Physics}\ }\textbf {\bibinfo {volume} {5}},\
  \bibinfo {pages} {27} (\bibinfo {year} {2008})}\BibitemShut {NoStop}%
\bibitem [{\citenamefont {Kraus}(1983)}]{Kra83}%
  \BibitemOpen
  \bibfield  {author} {\bibinfo {author} {\bibfnamefont {K.}~\bibnamefont
  {Kraus}},\ }\href@noop {} {\emph {\bibinfo {title} {States, effects and
  Operations. Fundamental notions of quantum theory}}}\ (\bibinfo  {publisher}
  {Academic Press},\ \bibinfo {year} {1983})\BibitemShut {NoStop}%
\bibitem [{\citenamefont {Aharonov}\ \emph {et~al.}(1988)\citenamefont
  {Aharonov}, \citenamefont {Albert},\ and\ \citenamefont {Vaidman}}]{Aha88}%
  \BibitemOpen
  \bibfield  {author} {\bibinfo {author} {\bibfnamefont {Y.}~\bibnamefont
  {Aharonov}}, \bibinfo {author} {\bibfnamefont {D.~Z.}\ \bibnamefont
  {Albert}}, \ and\ \bibinfo {author} {\bibfnamefont {L.}~\bibnamefont
  {Vaidman}},\ }\href@noop {} {\bibfield  {journal} {\bibinfo  {journal} {Phys.
  Rev. Lett.}\ }\textbf {\bibinfo {volume} {60}},\ \bibinfo {pages} {1351}
  (\bibinfo {year} {1988})}\BibitemShut {NoStop}%
\bibitem [{\citenamefont {Lanyon}\ \emph {et~al.}(2008)\citenamefont {Lanyon},
  \citenamefont {Barbieri}, \citenamefont {Almeida}, \citenamefont {Jennewein},
  \citenamefont {Ralph}, \citenamefont {Resch}, \citenamefont {Pryde},
  \citenamefont {O'Brien}, \citenamefont {Gilchrist},\ and\ \citenamefont
  {White}}]{Lan08}%
  \BibitemOpen
  \bibfield  {author} {\bibinfo {author} {\bibfnamefont {B.~P.}\ \bibnamefont
  {Lanyon}}, \bibinfo {author} {\bibfnamefont {M.}~\bibnamefont {Barbieri}},
  \bibinfo {author} {\bibfnamefont {M.~P.}\ \bibnamefont {Almeida}}, \bibinfo
  {author} {\bibfnamefont {T.}~\bibnamefont {Jennewein}}, \bibinfo {author}
  {\bibfnamefont {T.~C.}\ \bibnamefont {Ralph}}, \bibinfo {author}
  {\bibfnamefont {K.~J.}\ \bibnamefont {Resch}}, \bibinfo {author}
  {\bibfnamefont {G.~J.}\ \bibnamefont {Pryde}}, \bibinfo {author}
  {\bibfnamefont {J.~L.}\ \bibnamefont {O'Brien}}, \bibinfo {author}
  {\bibfnamefont {A.}~\bibnamefont {Gilchrist}}, \ and\ \bibinfo {author}
  {\bibfnamefont {A.~G.}\ \bibnamefont {White}},\ }\href@noop {} {\bibfield
  {journal} {\bibinfo  {journal} {Nature Physics}\ }\textbf {\bibinfo {volume}
  {5}},\ \bibinfo {pages} {134} (\bibinfo {year} {2008})}\BibitemShut {NoStop}%
\bibitem [{\citenamefont {Lvovsky}\ \emph {et~al.}(2001)\citenamefont
  {Lvovsky}, \citenamefont {Hansen}, \citenamefont {Aichele}, \citenamefont
  {Benson}, \citenamefont {Mlynek},\ and\ \citenamefont {Schiller}}]{Lvo01}%
  \BibitemOpen
  \bibfield  {author} {\bibinfo {author} {\bibfnamefont {A.~I.}\ \bibnamefont
  {Lvovsky}}, \bibinfo {author} {\bibfnamefont {H.}~\bibnamefont {Hansen}},
  \bibinfo {author} {\bibfnamefont {T.}~\bibnamefont {Aichele}}, \bibinfo
  {author} {\bibfnamefont {O.}~\bibnamefont {Benson}}, \bibinfo {author}
  {\bibfnamefont {J.}~\bibnamefont {Mlynek}}, \ and\ \bibinfo {author}
  {\bibfnamefont {S.}~\bibnamefont {Schiller}},\ }\href@noop {} {\bibfield
  {journal} {\bibinfo  {journal} {Phys. Rev. Lett.}\ }\textbf {\bibinfo
  {volume} {87}},\ \bibinfo {pages} {050402} (\bibinfo {year}
  {2001})}\BibitemShut {NoStop}%
\bibitem [{\citenamefont {Wenger}\ \emph {et~al.}(2004)\citenamefont {Wenger},
  \citenamefont {Tualle-Bruori},\ and\ \citenamefont {Grangier}}]{Wen04}%
  \BibitemOpen
  \bibfield  {author} {\bibinfo {author} {\bibfnamefont {J.}~\bibnamefont
  {Wenger}}, \bibinfo {author} {\bibfnamefont {R.}~\bibnamefont
  {Tualle-Bruori}}, \ and\ \bibinfo {author} {\bibfnamefont {P.}~\bibnamefont
  {Grangier}},\ }\href@noop {} {\bibfield  {journal} {\bibinfo  {journal}
  {Phys. Rev. Lett.}\ }\textbf {\bibinfo {volume} {92}},\ \bibinfo {pages}
  {153601} (\bibinfo {year} {2004})}\BibitemShut {NoStop}%
\bibitem [{\citenamefont {Zavatta}\ \emph {et~al.}(2004)\citenamefont
  {Zavatta}, \citenamefont {Viciani},\ and\ \citenamefont {Bellini}}]{Zav04}%
  \BibitemOpen
  \bibfield  {author} {\bibinfo {author} {\bibfnamefont {A.}~\bibnamefont
  {Zavatta}}, \bibinfo {author} {\bibfnamefont {S.}~\bibnamefont {Viciani}}, \
  and\ \bibinfo {author} {\bibfnamefont {M.}~\bibnamefont {Bellini}},\
  }\href@noop {} {\bibfield  {journal} {\bibinfo  {journal} {Phys. Rev. A}\
  }\textbf {\bibinfo {volume} {70}},\ \bibinfo {pages} {053821} (\bibinfo
  {year} {2004})}\BibitemShut {NoStop}%
\bibitem [{\citenamefont {Ourjoumtsev}\ \emph
  {et~al.}(2006{\natexlab{a}})\citenamefont {Ourjoumtsev}, \citenamefont
  {Tualle-Brouri}, \citenamefont {Laurat}, ,\ and\ \citenamefont
  {Grangier}}]{Our06}%
  \BibitemOpen
  \bibfield  {author} {\bibinfo {author} {\bibfnamefont {A.}~\bibnamefont
  {Ourjoumtsev}}, \bibinfo {author} {\bibfnamefont {R.}~\bibnamefont
  {Tualle-Brouri}}, \bibinfo {author} {\bibfnamefont {J.}~\bibnamefont
  {Laurat}}, , \ and\ \bibinfo {author} {\bibfnamefont {P.}~\bibnamefont
  {Grangier}},\ }\href@noop {} {\bibfield  {journal} {\bibinfo  {journal}
  {Science}\ }\textbf {\bibinfo {volume} {312}},\ \bibinfo {pages} {83}
  (\bibinfo {year} {2006}{\natexlab{a}})}\BibitemShut {NoStop}%
\bibitem [{\citenamefont {Ourjoumtsev}\ \emph
  {et~al.}(2007{\natexlab{a}})\citenamefont {Ourjoumtsev}, \citenamefont
  {Jeong}, \citenamefont {Tualle-Brouri},\ and\ \citenamefont
  {Grangier}}]{Our07}%
  \BibitemOpen
  \bibfield  {author} {\bibinfo {author} {\bibfnamefont {A.}~\bibnamefont
  {Ourjoumtsev}}, \bibinfo {author} {\bibfnamefont {H.}~\bibnamefont {Jeong}},
  \bibinfo {author} {\bibfnamefont {R.}~\bibnamefont {Tualle-Brouri}}, \ and\
  \bibinfo {author} {\bibfnamefont {P.}~\bibnamefont {Grangier}},\ }\href@noop
  {} {\bibfield  {journal} {\bibinfo  {journal} {Nature}\ }\textbf {\bibinfo
  {volume} {448}},\ \bibinfo {pages} {784} (\bibinfo {year}
  {2007}{\natexlab{a}})}\BibitemShut {NoStop}%
\bibitem [{\citenamefont {Eisert}\ \emph {et~al.}(2002)\citenamefont {Eisert},
  \citenamefont {Scheel},\ and\ \citenamefont {Plenio}}]{Eis02}%
  \BibitemOpen
  \bibfield  {author} {\bibinfo {author} {\bibfnamefont {J.}~\bibnamefont
  {Eisert}}, \bibinfo {author} {\bibfnamefont {S.}~\bibnamefont {Scheel}}, \
  and\ \bibinfo {author} {\bibfnamefont {M.~B.}\ \bibnamefont {Plenio}},\
  }\href@noop {} {\bibfield  {journal} {\bibinfo  {journal} {Phys. Rev. Lett.}\
  }\textbf {\bibinfo {volume} {89}},\ \bibinfo {pages} {137903} (\bibinfo
  {year} {2002})}\BibitemShut {NoStop}%
\bibitem [{\citenamefont {Fiurasek}(2002)}]{Fiu02}%
  \BibitemOpen
  \bibfield  {author} {\bibinfo {author} {\bibfnamefont {J.}~\bibnamefont
  {Fiurasek}},\ }\href@noop {} {\bibfield  {journal} {\bibinfo  {journal}
  {Phys. Rev. Lett.}\ }\textbf {\bibinfo {volume} {89}},\ \bibinfo {pages}
  {137904} (\bibinfo {year} {2002})}\BibitemShut {NoStop}%
\bibitem [{\citenamefont {Niset}\ \emph {et~al.}(2009)\citenamefont {Niset},
  \citenamefont {Fiurasek},\ and\ \citenamefont {Cerf}}]{Nis09}%
  \BibitemOpen
  \bibfield  {author} {\bibinfo {author} {\bibfnamefont {J.}~\bibnamefont
  {Niset}}, \bibinfo {author} {\bibfnamefont {J.}~\bibnamefont {Fiurasek}}, \
  and\ \bibinfo {author} {\bibfnamefont {N.~J.}\ \bibnamefont {Cerf}},\
  }\href@noop {} {\bibfield  {journal} {\bibinfo  {journal} {Phys. Rev. Lett.}\
  }\textbf {\bibinfo {volume} {102}},\ \bibinfo {pages} {120501} (\bibinfo
  {year} {2009})}\BibitemShut {NoStop}%
\bibitem [{\citenamefont {Ourjoumtsev}\ \emph
  {et~al.}(2007{\natexlab{b}})\citenamefont {Ourjoumtsev}, \citenamefont
  {Dantan}, \citenamefont {Tualle-Brouri},\ and\ \citenamefont
  {Grangier}}]{Our07a}%
  \BibitemOpen
  \bibfield  {author} {\bibinfo {author} {\bibfnamefont {A.}~\bibnamefont
  {Ourjoumtsev}}, \bibinfo {author} {\bibfnamefont {A.}~\bibnamefont {Dantan}},
  \bibinfo {author} {\bibfnamefont {R.}~\bibnamefont {Tualle-Brouri}}, \ and\
  \bibinfo {author} {\bibfnamefont {P.}~\bibnamefont {Grangier}},\ }\href@noop
  {} {\bibfield  {journal} {\bibinfo  {journal} {Phys. Rev. Lett.}\ }\textbf
  {\bibinfo {volume} {98}},\ \bibinfo {pages} {030502} (\bibinfo {year}
  {2007}{\natexlab{b}})}\BibitemShut {NoStop}%
\bibitem [{\citenamefont {Takahashi}\ \emph {et~al.}(2010)\citenamefont
  {Takahashi}, \citenamefont {Neergaard-Nielsen}, \citenamefont {Takeuchi},
  \citenamefont {Takeoka}, \citenamefont {Hayasaka}, \citenamefont {Furusawa},\
  and\ \citenamefont {Sasaki}}]{Tak10}%
  \BibitemOpen
  \bibfield  {author} {\bibinfo {author} {\bibfnamefont {H.}~\bibnamefont
  {Takahashi}}, \bibinfo {author} {\bibfnamefont {J.~S.}\ \bibnamefont
  {Neergaard-Nielsen}}, \bibinfo {author} {\bibfnamefont {M.}~\bibnamefont
  {Takeuchi}}, \bibinfo {author} {\bibfnamefont {M.}~\bibnamefont {Takeoka}},
  \bibinfo {author} {\bibfnamefont {K.}~\bibnamefont {Hayasaka}}, \bibinfo
  {author} {\bibfnamefont {A.}~\bibnamefont {Furusawa}}, \ and\ \bibinfo
  {author} {\bibfnamefont {M.}~\bibnamefont {Sasaki}},\ }\href@noop {}
  {\bibfield  {journal} {\bibinfo  {journal} {Nature Photonics}\ }\textbf
  {\bibinfo {volume} {4}},\ \bibinfo {pages} {178} (\bibinfo {year}
  {2010})}\BibitemShut {NoStop}%
\bibitem [{\citenamefont {Ferreyrol}\ \emph {et~al.}(2010)\citenamefont
  {Ferreyrol}, \citenamefont {Barbieri}, \citenamefont {Blandino},
  \citenamefont {Fossier}, \citenamefont {Tualle-Brouri},\ and\ \citenamefont
  {Grangier}}]{Fer10}%
  \BibitemOpen
  \bibfield  {author} {\bibinfo {author} {\bibfnamefont {F.}~\bibnamefont
  {Ferreyrol}}, \bibinfo {author} {\bibfnamefont {M.}~\bibnamefont {Barbieri}},
  \bibinfo {author} {\bibfnamefont {R.}~\bibnamefont {Blandino}}, \bibinfo
  {author} {\bibfnamefont {S.}~\bibnamefont {Fossier}}, \bibinfo {author}
  {\bibfnamefont {R.}~\bibnamefont {Tualle-Brouri}}, \ and\ \bibinfo {author}
  {\bibfnamefont {P.}~\bibnamefont {Grangier}},\ }\href@noop {} {\bibfield
  {journal} {\bibinfo  {journal} {Phys. Rev. Lett.}\ }\textbf {\bibinfo
  {volume} {104}},\ \bibinfo {pages} {123603} (\bibinfo {year}
  {2010})}\BibitemShut {NoStop}%
\bibitem [{\citenamefont {Xiang}\ \emph {et~al.}(2010)\citenamefont {Xiang},
  \citenamefont {Ralph}, \citenamefont {Lund}, \citenamefont {Walk},\ and\
  \citenamefont {Pryde}}]{Xia10}%
  \BibitemOpen
  \bibfield  {author} {\bibinfo {author} {\bibfnamefont {G.~Y.}\ \bibnamefont
  {Xiang}}, \bibinfo {author} {\bibfnamefont {T.~C.}\ \bibnamefont {Ralph}},
  \bibinfo {author} {\bibfnamefont {A.~P.}\ \bibnamefont {Lund}}, \bibinfo
  {author} {\bibfnamefont {N.}~\bibnamefont {Walk}}, \ and\ \bibinfo {author}
  {\bibfnamefont {G.~J.}\ \bibnamefont {Pryde}},\ }\href@noop {} {\bibfield
  {journal} {\bibinfo  {journal} {Nature Photonics}\ }\textbf {\bibinfo
  {volume} {4}},\ \bibinfo {pages} {316} (\bibinfo {year} {2010})}\BibitemShut
  {NoStop}%
\bibitem [{\citenamefont {Kiesel}\ \emph {et~al.}(2005)\citenamefont {Kiesel},
  \citenamefont {Schmid}, \citenamefont {Weber}, \citenamefont {Ursin},\ and\
  \citenamefont {Weinfurter}}]{Kie05}%
  \BibitemOpen
  \bibfield  {author} {\bibinfo {author} {\bibfnamefont {N.}~\bibnamefont
  {Kiesel}}, \bibinfo {author} {\bibfnamefont {C.}~\bibnamefont {Schmid}},
  \bibinfo {author} {\bibfnamefont {U.}~\bibnamefont {Weber}}, \bibinfo
  {author} {\bibfnamefont {R.}~\bibnamefont {Ursin}}, \ and\ \bibinfo {author}
  {\bibfnamefont {H.}~\bibnamefont {Weinfurter}},\ }\href@noop {} {\bibfield
  {journal} {\bibinfo  {journal} {Phys. Rev. Lett.}\ }\textbf {\bibinfo
  {volume} {95}},\ \bibinfo {pages} {210505} (\bibinfo {year}
  {2005})}\BibitemShut {NoStop}%
\bibitem [{\citenamefont {Bongioanni}\ \emph {et~al.}(2010)\citenamefont
  {Bongioanni}, \citenamefont {Sansoni}, \citenamefont {Sciarrino},
  \citenamefont {Vallone},\ and\ \citenamefont {Mataloni}}]{Bon10}%
  \BibitemOpen
  \bibfield  {author} {\bibinfo {author} {\bibfnamefont {I.}~\bibnamefont
  {Bongioanni}}, \bibinfo {author} {\bibfnamefont {L.}~\bibnamefont {Sansoni}},
  \bibinfo {author} {\bibfnamefont {F.}~\bibnamefont {Sciarrino}}, \bibinfo
  {author} {\bibfnamefont {G.}~\bibnamefont {Vallone}}, \ and\ \bibinfo
  {author} {\bibfnamefont {P.}~\bibnamefont {Mataloni}},\ }\href@noop {}
  {\bibfield  {journal} {\bibinfo  {journal} {Phys. Rev. A}\ }\textbf {\bibinfo
  {volume} {82}},\ \bibinfo {pages} {042307} (\bibinfo {year}
  {2010})}\BibitemShut {NoStop}%
\bibitem [{\citenamefont {Barbieri}\ \emph {et~al.}(2010)\citenamefont
  {Barbieri}, \citenamefont {Spagnolo}, \citenamefont {Genoni}, \citenamefont
  {Ferreyrol}, \citenamefont {Blandino}, \citenamefont {Paris}, \citenamefont
  {Grangier},\ and\ \citenamefont {Tualle-Brouri}}]{Bar10}%
  \BibitemOpen
  \bibfield  {author} {\bibinfo {author} {\bibfnamefont {M.}~\bibnamefont
  {Barbieri}}, \bibinfo {author} {\bibfnamefont {N.}~\bibnamefont {Spagnolo}},
  \bibinfo {author} {\bibfnamefont {M.~G.}\ \bibnamefont {Genoni}}, \bibinfo
  {author} {\bibfnamefont {F.}~\bibnamefont {Ferreyrol}}, \bibinfo {author}
  {\bibfnamefont {R.}~\bibnamefont {Blandino}}, \bibinfo {author}
  {\bibfnamefont {M.~G.~A.}\ \bibnamefont {Paris}}, \bibinfo {author}
  {\bibfnamefont {P.}~\bibnamefont {Grangier}}, \ and\ \bibinfo {author}
  {\bibfnamefont {R.}~\bibnamefont {Tualle-Brouri}},\ }\href@noop {} {\bibfield
   {journal} {\bibinfo  {journal} {Phys. Rev. A}\ }\textbf {\bibinfo {volume}
  {82}},\ \bibinfo {pages} {063833} (\bibinfo {year} {2010})}\BibitemShut
  {NoStop}%
\bibitem [{\citenamefont {Nielsen}\ and\ \citenamefont
  {Chuang}(2000)}]{Nielsen_Chuang}%
  \BibitemOpen
  \bibfield  {author} {\bibinfo {author} {\bibfnamefont {M.}~\bibnamefont
  {Nielsen}}\ and\ \bibinfo {author} {\bibfnamefont {I.}~\bibnamefont
  {Chuang}},\ }\href@noop {} {\emph {\bibinfo {title} {Quantum Computation and
  Quantum Information}}},\ Cambridge Series on Information and the Natural
  Sciences\ (\bibinfo  {publisher} {Cambridge University Press},\ \bibinfo
  {year} {2000})\BibitemShut {NoStop}%
\bibitem [{\citenamefont {Berry}\ \emph {et~al.}(1979)\citenamefont {Berry},
  \citenamefont {Balazs}, \citenamefont {Tabor},\ and\ \citenamefont
  {Voros}}]{Ber79}%
  \BibitemOpen
  \bibfield  {author} {\bibinfo {author} {\bibfnamefont {M.}~\bibnamefont
  {Berry}}, \bibinfo {author} {\bibfnamefont {N.}~\bibnamefont {Balazs}},
  \bibinfo {author} {\bibfnamefont {M.}~\bibnamefont {Tabor}}, \ and\ \bibinfo
  {author} {\bibfnamefont {A.}~\bibnamefont {Voros}},\ }\href@noop {}
  {\bibfield  {journal} {\bibinfo  {journal} {Ann. Phys}\ }\textbf {\bibinfo
  {volume} {122}},\ \bibinfo {pages} {26} (\bibinfo {year} {1979})}\BibitemShut
  {NoStop}%
\bibitem [{\citenamefont {Cohen-Tannoudji}()}]{Coh83}%
  \BibitemOpen
  \bibfield  {author} {\bibinfo {author} {\bibfnamefont {C.}~\bibnamefont
  {Cohen-Tannoudji}},\ }\href@noop {} {}\bibinfo {note} {Lecture notes of
  Coll\`ege de France, available at:
  http://www.phys.ens.fr/cours/college-de-france/1983-84/1983-84.htm}\BibitemShut
  {NoStop}%
\bibitem [{\citenamefont {Ralph}\ and\ \citenamefont {Lund}(2009)}]{Tim08}%
  \BibitemOpen
  \bibfield  {author} {\bibinfo {author} {\bibfnamefont {T.~C.}\ \bibnamefont
  {Ralph}}\ and\ \bibinfo {author} {\bibfnamefont {A.~P.}\ \bibnamefont
  {Lund}},\ }in\ \href@noop {} {\emph {\bibinfo {booktitle} {Quantum
  Communication Measurement and Computing Proceedings of 9th International
  Conference}}},\ Vol.\ \bibinfo {volume} {1110},\ \bibinfo {editor} {edited
  by\ \bibinfo {editor} {\bibfnamefont {A.~I.}\ \bibnamefont {Lvovsky}}}\
  (\bibinfo  {publisher} {AIP Conference proceedings},\ \bibinfo {year}
  {2009})\ p.\ \bibinfo {pages} {155}\BibitemShut {NoStop}%
\bibitem [{\citenamefont {Ourjoumtsev}\ \emph
  {et~al.}(2006{\natexlab{b}})\citenamefont {Ourjoumtsev}, \citenamefont
  {Tualle-Brouri},\ and\ \citenamefont {Grangier}}]{Our06a}%
  \BibitemOpen
  \bibfield  {author} {\bibinfo {author} {\bibfnamefont {A.}~\bibnamefont
  {Ourjoumtsev}}, \bibinfo {author} {\bibfnamefont {R.}~\bibnamefont
  {Tualle-Brouri}}, \ and\ \bibinfo {author} {\bibfnamefont {P.}~\bibnamefont
  {Grangier}},\ }\href@noop {} {\bibfield  {journal} {\bibinfo  {journal}
  {Phys. Rev. Lett.}\ }\textbf {\bibinfo {volume} {96}},\ \bibinfo {pages}
  {213601} (\bibinfo {year} {2006}{\natexlab{b}})}\BibitemShut {NoStop}%
\bibitem [{\citenamefont {Ferreyrol}\ \emph {et~al.}(2011)\citenamefont
  {Ferreyrol}, \citenamefont {Blandino}, \citenamefont {Barbieri},
  \citenamefont {Tualle-Brouri},\ and\ \citenamefont {Grangier}}]{Fer10a}%
  \BibitemOpen
  \bibfield  {author} {\bibinfo {author} {\bibfnamefont {F.}~\bibnamefont
  {Ferreyrol}}, \bibinfo {author} {\bibfnamefont {R.}~\bibnamefont {Blandino}},
  \bibinfo {author} {\bibfnamefont {M.}~\bibnamefont {Barbieri}}, \bibinfo
  {author} {\bibfnamefont {R.}~\bibnamefont {Tualle-Brouri}}, \ and\ \bibinfo
  {author} {\bibfnamefont {P.}~\bibnamefont {Grangier}},\ }\href@noop {}
  {\bibfield  {journal} {\bibinfo  {journal} {Phys. Rev. A}\ }\textbf {\bibinfo
  {volume} {83}},\ \bibinfo {pages} {063801} (\bibinfo {year}
  {2011})}\BibitemShut {NoStop}%
\bibitem [{\citenamefont {Parigi}\ \emph {et~al.}(2007)\citenamefont {Parigi},
  \citenamefont {Zavatta}, \citenamefont {Kim},\ and\ \citenamefont
  {Bellini}}]{Par07}%
  \BibitemOpen
  \bibfield  {author} {\bibinfo {author} {\bibfnamefont {V.}~\bibnamefont
  {Parigi}}, \bibinfo {author} {\bibfnamefont {A.}~\bibnamefont {Zavatta}},
  \bibinfo {author} {\bibfnamefont {M.~S.}\ \bibnamefont {Kim}}, \ and\
  \bibinfo {author} {\bibfnamefont {M.}~\bibnamefont {Bellini}},\ }\href@noop
  {} {\bibfield  {journal} {\bibinfo  {journal} {Science}\ }\textbf {\bibinfo
  {volume} {317}},\ \bibinfo {pages} {1890} (\bibinfo {year}
  {2007})}\BibitemShut {NoStop}%
\bibitem [{\citenamefont {Zavatta}\ \emph {et~al.}(2009)\citenamefont
  {Zavatta}, \citenamefont {Parigi}, \citenamefont {Kim}, \citenamefont
  {Jeong},\ and\ \citenamefont {Bellini}}]{Zav09}%
  \BibitemOpen
  \bibfield  {author} {\bibinfo {author} {\bibfnamefont {A.}~\bibnamefont
  {Zavatta}}, \bibinfo {author} {\bibfnamefont {V.}~\bibnamefont {Parigi}},
  \bibinfo {author} {\bibfnamefont {M.~S.}\ \bibnamefont {Kim}}, \bibinfo
  {author} {\bibfnamefont {H.}~\bibnamefont {Jeong}}, \ and\ \bibinfo {author}
  {\bibfnamefont {M.}~\bibnamefont {Bellini}},\ }\href@noop {} {\bibfield
  {journal} {\bibinfo  {journal} {Phys. Rev. Lett.}\ }\textbf {\bibinfo
  {volume} {103}},\ \bibinfo {pages} {140406} (\bibinfo {year}
  {2009})}\BibitemShut {NoStop}%
\bibitem [{\citenamefont {Zavatta}\ \emph {et~al.}(2007)\citenamefont
  {Zavatta}, \citenamefont {Parigi},\ and\ \citenamefont {Bellini}}]{Zav07}%
  \BibitemOpen
  \bibfield  {author} {\bibinfo {author} {\bibfnamefont {A.}~\bibnamefont
  {Zavatta}}, \bibinfo {author} {\bibfnamefont {V.}~\bibnamefont {Parigi}}, \
  and\ \bibinfo {author} {\bibfnamefont {M.}~\bibnamefont {Bellini}},\
  }\href@noop {} {\bibfield  {journal} {\bibinfo  {journal} {Phys. Rev. A}\
  }\textbf {\bibinfo {volume} {75}},\ \bibinfo {pages} {052106} (\bibinfo
  {year} {2007})}\BibitemShut {NoStop}%
\end{thebibliography}%
\bibliographystyle{apsrev4-1}

\end{document}